\documentclass[submission, Phys]{SciPost}

\usepackage{hyperref} 
\usepackage{graphicx}
\usepackage{tikz}
\usepackage{epstopdf}
\usepackage{mathtools}
\usepackage{dcolumn}
\usepackage{bm}
\usepackage{amsmath}
\usepackage{mathrsfs}
\usepackage{siunitx}
\usepackage{booktabs}
\usepackage{multirow}
\usepackage{soul}
\usepackage{lineno}

\usepackage{xcolor} 
\ifx\nocolor\undefined 
  \newcommand{\ttred}[1]{\textcolor{black}{#1}} 
  
  \newcommand{\tblue}[1]{\textcolor{black}{#1}} 
  \newcommand{\tred}[1]{\textcolor{black}{#1}} 
  \newcommand{\tcyan}[1]{\textcolor{black}{#1}} 
  \newcommand{\tgreen}[1]{\textcolor{black}{#1}} 
  \newcommand{\tteal}[1]{\textcolor{black}{#1}} 
  \newcommand{\tolive}[1]{\textcolor{black}{#1}} 
  \newcommand{\tviolet}[1]{\textcolor{black}{#1}} 
  \newcommand{\torange}[1]{\textcolor{black}{#1}} 
  \newcommand{\strout}[1]{\sout{#1}}
\else
   \newcommand{\tblue}[1]{{#1}}
   \newcommand{\tred}[1]{{#1}}
   \newcommand{\tcyan}[1]{{#1}}
   \newcommand{\tgreen}[1]{{#1}}
   \newcommand{\strout}[1]{}
\fi

\makeatletter
\renewcommand\st[1]{\@bsphack\@esphack}
\makeatother
\renewcommand{\hl}[1]{#1}

\begin{document}

\begin{center}{\Large \textbf{Pressure-induced transitions in FePS$_3$: 
Structural, magnetic and electronic properties
}}\end{center}

\begin{center}
Shiyu Deng\textsuperscript{1*},
Siyu Chen\textsuperscript{1},
Bartomeu Monserrat\textsuperscript{1,2},
Emilio Artacho\textsuperscript{1,3,4*},
Siddharth S Saxena\textsuperscript{1*}
\end{center}

\begin{center}
{\bf 1} Cavendish Laboratory, University of Cambridge, J. J. Thomson Avenue, Cambridge, CB3 0HE, United Kingdom
\\
{\bf 2} Department of Materials Science and Metallurgy, University of Cambridge, 27 Charles Babbage Road, Cambridge CB3 0FS, United Kingdom
\\
{\bf 3} CIC Nanogune BRTA and DIPC, Tolosa Hiribidea 76, 20018 San Sebastian, Spain
\\
{\bf 4} Ikerbasque, Basque Foundation for Science, 48011 Bilbao, Spain
\\
* sd864@cam.ac.uk, e.artacho@nanogune.eu, sss21@cam.ac.uk
\end{center}

\begin{center}
\today
\end{center}


\section*{Abstract}
{\bf
FePS$_3$ is a prototype van der Waals layered antiferromagnet and a Mott insulator under ambient conditions, 
which has been recently reported to go through a pressure-induced dimensionality crossover and an insulator-to-metal transition.
These transitions also lead to the appearance of a novel magnetic metallic state.
To further understand these emergent structural and physical properties,
we have performed a first-principles study using van der Waals and Hubbard $U$ corrected density functional theory including a random structure search.
Our computational study attempts to interpret the experimental coexistence of the low- and intermediate-pressure phases and we predict a novel high-pressure phase with distinctive dimensionality and different possible origins of metallicity.
}

\vspace{10pt}
\noindent\rule{\textwidth}{1pt}
\tableofcontents\thispagestyle{fancy}
\noindent\rule{\textwidth}{1pt}
\vspace{10pt}

\section{Introduction}

\tblue{
Two-dimensional (2D) materials with tunable electrical and magnetic properties 
have substantial potential in the design of 
atomically thin 
devices for data storage, quantum computing\cite{Gong2019_2DMagnet_Review} \torange{and clean energy generation related with photocatalytic water splitting applications \cite{2023_FePX3_water_splitting_application}}. 
\tcyan{For several decades the \tgreen{study} of 2D phenomena in magnets has been inhibited 
due to particular interpretations of the well-known Hohenberg-Mermin-Wagner theorem \cite{1966_Mermin_Wagner_Theorem,1967_Hobenberg}. 
However, this is beginning to change given the recent} observation of intrinsic long-range magnetic order in 2D materials, including the ferromagnetic (FM) conductor Fe$_3$GeTe$_2$ \cite{2016_Fe3GeTe2}, 
magnetic insulator Cr$_2$Ge$_2$Te$_6$ \cite{Gong2017_Cr2Ge2Te6} 
and antiferromagnetic (AFM) insulators $TM\rm PX_3$ ($TM$ = Mn, Fe, Ni, V, etc., and $X$ = S, Se) \cite{LeFlem1982_TMPS3_OldTheoModel, 2006_Wildes_MnPS3,2016_Lee_IsingTypeOrder_FePS3}
\tcyan{which are the focus of this study}.
It has been found that magnetic anisotropy could open up an excitation energy gap 
to counteract the enhanced thermal fluctuations in low dimensional materials \cite{Halperin2019_HMW_theorem}. 
\tolive{Different from} conventional thin films \cite{martin2016thin}, 
2D materials \tcyan{bound only by} \tolive{van der Waals} (vdW) interactions offer more precise manufacturing
control and reproducibility \cite{2022_ProximityEffect_Monolayer_MPX3,2022_Zhou_SpinShearCoupling_vdWAFM}.
}

\tblue{
\tcyan{A recently realized advantage of 2D magnets is that they are often materials 
where the ground state is dominated by the physics of strong correlations.
These materials thus 
become fertile ground for exploring novel phases and emergent phenomena.}
For instance, the well-known high-$T_c$ cuprates are obtained via doping the 2D AFM precursors \cite{1986_First_Cuprate,1988_Cuprate_120K}. 
Most recently, the pressure-induced superconductivity (SC) \torange{in the vicinity of AFM order has attracted much interest, with prominent examples being the iron-based layered compound LaFeAsO \cite{2014_kumar_pressure_SC_LaFeAsO}, FeSe \cite{2009_Imai_FeSe_SC_pressure,2021_PNAS_pressure_highT_SC_FeSe}, and other transition-metal compounds e.g. MnSe \cite{2021_NatComm_pressure_SC_MnSe}, CrAs \cite{2014_NatComm_SC_near_AFM_CrAs} and AuTe$_2$Br \cite{2022_Pressure_induced_SC_AuTe2Br}.} 
\tcyan{In such systems, }
reduced dimensionality is believed to enhance SC \cite{2001_Monthoux_SC_2D_3D,2020_NatPhys_SC_LowD}. 
However, 
\tcyan{the understanding of the fundamental physics of these emergent phenomena is still \tolive{hindering} proper theoretical formulation.
In this regard, the $TM\rm{P}X_3$ family represents a promising new avenue of research.}
}

\tblue{
Transition metal phosphorous trichalcogenides $TM\rm{P}X_3$
have proven to be ideal examples 
\tcyan{of exploration through tuning of}
the structural, magnetic and electronic properties in 2D vdW layered systems \tcyan{exhibiting} both long-range magnetic order and strong correlations \cite{2016_2DMagnet_outlook, 2018_Burch_Magnetism_vdW_Materials}.} 
Being Mott or charge-transfer insulators at ambient pressure, 
the band gap of $TM\rm{P}X_3$ could be tuned 
\tcyan{systematically via \tolive{the} application of pressure.} 
Recent high-pressure studies have revealed spin-crossover transitions, insulator-to-metal transitions (IMT) and even the emergence of SC in these compounds \cite{2016Wang_MnPX3_spinCrossover, 2019_MnPX3_NiPX3_pressure_IMT,2018_Haines_highPressure_XRD_FePS3,2018_Wang_pressure_SC_FePX3,2020_Coak_TuneDimensionality_TMPS3,2022_Raman_DFT_FePS3}.
\tblue{
\tcyan{In this family, }
FePS$_3$ is \tcyan{an ideal} prototype to start with, 
considering the Ising nature of the magnetic moments \tolive{of} Fe ions \cite{2002_Rule_Contrasting_AFMorder_FePS3_MnPS3,2016_Lancon_FePS3_MagStructure}, 
small band gap of about $1.5$\,eV, 
lowest resistivity of $1.0\times10^{12}$\,$\Omega$\,cm \cite{1979_TMPX3_resistivity_bandGap} at ambient pressure 
and experimental evidence \tcyan{for \tolive{the} evolution of} the structural, electronic and magnetic properties at high pressures \cite{2018_Haines_highPressure_XRD_FePS3,2018_Wang_pressure_SC_FePX3,2020_Coak_TuneDimensionality_TMPS3,2022_Raman_DFT_FePS3}.}

\tblue{
For FePS$_3$, it has been challenging} 
\tcyan{to accurately ascertain \tolive{its} crystallographic and electronic properties 
due to the lack of precise atomic positions from the experimental data. 
More generally, correlation effects \tred{always represent a challenge for 
computational simulations.}}
\tteal{While experimentally observed spin-spin short-range correlations are hard to tackle with first-principles calculations, it would be worthwhile to carefully
examine these effects, particularly near the transition from 2D vdW compounds to more 3D bonded configurations.}

In 2018, Haines $et\ al.$ \cite{2018_Haines_highPressure_XRD_FePS3} and Wang $et\ al.$ \cite{2018_Wang_pressure_SC_FePX3} performed independent \tolive{high} pressure experiments on FePS$_3$ powder samples. 
Both groups observed the IMT and volume collapse in response to the external pressure but proposed incompatible models for the high-pressure (HP) phase. 
Wang $et\ al.$ \cite{2018_Wang_pressure_SC_FePX3} claimed that the \tblue{low-pressure (LP)} monoclinic symmetry remains until the \tblue{HP} region and the in-plane lattice collapse contributed the most to the volume collapse during the iso-structural phase transition at $\sim$ 13 GPa. 
They also reported that when the HP phase turns into a metallic phase in the case of FePSe$_3$, which is a related compound with similar structural and magnetic properties, 
\torange{SC was observed} at 2.5 K and 9.0 GPa \cite{2018_Wang_pressure_SC_FePX3}.
Meanwhile, Haines $et\ al.$ \cite{2018_Haines_highPressure_XRD_FePS3} claimed that there are two transitions. 
The first \tolive{happens} around 4 GPa via inter-planar sliding.  
The LP \tolive{phase} evolves into HP-I without symmetry and dimensionality change 
and remains insulating. 
The next occurs around 14 GPa with an interlayer lattice collapse and the bulk symmetry changed to $P\overline{3}1m$ \tolive{in HP-II}. 
The HP-II \tolive{phase} was determined to be metallic and more 3D-like. 
Subsequently, two computational \tteal{studies}
by Zheng $et\ al.$ \cite{Zheng2019_abinito} and Evarestov $et\ al.$ \cite{2020_Evarestov_DFT_FePX3} 
did not come up with consistent conclusions 
\tteal{regarding the origins of IMT or the impact of magnetic configurations on the crystal structure.}
\ttred{A later Raman spectroscopy work by Das $et\ al.$ has detected two phase transitions at 4.6 GPa and 12.0 GPa, complementing the previous experimental observations and models \cite{2022_Raman_DFT_FePS3}.
}
The most recent study by Jarvis $et\ al.$ \cite{2023_Jarvis_comparative_single_powder_FePS3} compares the different experiments utilizing powder samples with and without the helium pressure medium, and includes the single crystal diffraction results in the discussions. 
The experimental environment plays an essential role in the high-pressure behaviour.

The detailed magnetic structure of FePS$_3$ is challenging to resolve
\torange{even at ambient pressure \cite{2002_Rule_Contrasting_AFMorder_FePS3_MnPS3,2007_Rule_neutron_MagneticStructure_FePS3,2020_Wildes_Biquadratic_FePS3,2022_Zeng_FePS3_MagTran_DFT} and affects the high-pressure phases as well}.
A recent \tteal{study} by Coak $et\ al.$ \cite{2021_Coak_MagPhase_FePS3} further 
\tred{characterised} the LP, HP-I and HP-II phases \tolive{of} FePS$_3$ up to about 18 GPa. 
They first examined the evolution of magnetic phases with pressure using powder neutron diffraction, and 
proposed that from LP to HP-I the interlayer interaction transforms from antiferromagnetic to ferromagnetic. 
In the metallic HP-II phase, the long-range magnetic order is suppressed while a form of short-range order emerges \cite{2021_Coak_MagPhase_FePS3}. 
This is \tteal{a rather different result from} the spin crossover transition model proposed by Wang $et\ al.$ \cite{2018_Wang_pressure_SC_FePX3}, where the final state \tolive{is} claimed to be non-polarised. 
\torange{The magnetic interactions in $TM$P$X_3$ are complex and offer further opportunities to explore the underlying physics \cite{2000_Ordering_SpinFlop_MnPS3,2022_FM_monolayer_MP3,2023_Field_induced_Spin_Reorientation_MnPS3}.}

To \tteal{understand} how external pressure tunes the dimensionality, structural, electronic and magnetic properties in FePS$_3$, 
\tteal{it is essential to grasp the systematic evolution of the crystalline phase with applied pressure.}
Given the fact that 
\tred{different} experimental environments and setup\tgreen{s} 
\tred{affect} the high-pressure behaviour, we re-examine the high-pressure structures via a random structure search \tred{method using density functional theory}. 
The advantage \tteal{of this method} is that it does not require empirical knowledge of the \tteal{experimental findings} and thus allows us to search for the most stable and metastable structures from the energy \tteal{considerations}. 
We are able to reproduce the proposed phases and predict a novel one in the high-pressure region. 
Moreover, we look into the interlayer sliding and dimensionality change in FePS$_3$ \tteal{as it undergoes the} phase transitions in detail, together with the evolution of electronic and magnetic properties.


\section{Methods}

First principles calculations are performed based on 
density functional theory (DFT), using the Cambridge Serial Total Energy Package ({\sc CASTEP}) code \cite{2000_CASTEP,2002_CASTEP}, \tblue{version 19.1}.
The generalized-gradient-approximation (GGA) within the framework of Perdew, Burke, and Ernzerhof (PBE) \cite{Perdew1996GeneralizedSimple} is used for the exchange-correlation functional. 
\tblue{We utilize the “on-the-fly”\cite{pickard2006fly} generated ultrasoft 
pseudopotentials based on the formalism of Vanderbilt \cite{1990_OTFP}, as implemented in {\sc CASTEP}. 
The error achieved by this set is $0.4$\,meV/atom within the test framework of Lejaeghere 
$et\ al$ \cite{2014_DFT_error_estimate}.}

The Broyden-Fletcher-Goldfarb-Shanno (BFGS) algorithm \cite{1985_BFGS} was used for the geometry optimization, 
\tblue{with the force convergence tolerance being $0.05$\,eV/\AA}. 
The vdW interactions have been taken into account using the Tkatchenko-Scheffler approach \cite{2009_TS_correction}. 
Correlation effects have also been considered within the framework of the DFT+$U$ method \cite{1963_Hubbard_elelctron_correlations,1991_DFT+U,1995_MottHubbard_Insulators}, \tred{allowing for spin polarization. 
\ttred{Different starting spin configurations} of the self-consistency (SCF) cycle allow for different \ttred{spin polarization results and the local spin arrangements among Fe atoms. 
Here,} both the SCF and structural relaxations \ttred{are constrained in the relative spin orientations, yet} without constraint on the magnitude of local spins, 
allowing for magnitude variations on prescribed spin arrangements (using {\sc CASTEP}'s $\tt SPIN\_FIX$ and $\tt GEOM\_SPIN\_FIX$ option for SCF and relaxations)
that are later compared in energy.}
For the localized $d$ orbitals of the Fe atoms a Hubbard effective electron-electron repulsion $U$ is taken to be $2.5$\,eV, \tred{as used in similar studies \cite{2008_lebegue_DFT+U_Fe,Zheng2019_abinito,2020_Evarestov_DFT_FePX3}}. 
\tolive{The} effect of varying this parameter 
in the results is included in the results and discussion below.

The convergence criterion for the overall formation energy has been set as $10^{-6}$ eV/atom for all enthalpy calculations. 
To ensure proper convergence within reasonable computational time, we perform a series of tests running the self-consistent single-point simulations.
\tred{Tests are shown in Appendix~\ref{app:tests}}.
Valence-electron wave-functions are described with a plane-wave basis set, 
with a cutoff energy for the expansion of $550$\,eV, \tred{as selected from the tests in \tteal{Appendix \ref{app:tests}}}.
The sampling of discretized $\textbf{\textit{k}}$-points across the Brillouin zone follows the Monkhorst-Pack scheme \cite{1977_MP_kmesh} and is determined to be $0.03$\,\AA$^{-1}$ along each axis ({\sc CASTEP}'s $\tt KPOINTS\_MP\_SPACING$). 
The Fourier transform grid for the electron density is larger than that of the wave functions by a factor of 2.0. 

To search for the stable and metastable structures of FePS$_3$ at high pressures, we perform an $ab\ initio$ random structure search (AIRSS) \cite{Pickard2011AbSearching} \ttred{to generate random structures that are fully relaxed with} CASTEP \cite{2000_CASTEP,2002_CASTEP}. 
We carry out the AIRSS structure search at 0, 10 and 20 GPa using randomly generated structures containing between 1 and 4 chemical formula units of FePS$_3$ within one simulation cell. 
A random set of unit cell lengths and angles \tolive{is} generated as the cell volume is re-normalized to a random value within $\pm$50\% of the volume derived from the known structures. 
\tred{Atomic positions are also selected at random within the cell, 
\tblue{with the constraint of the minimum distance being 3.3 \AA\ for Fe-Fe, Fe-P and S-S, \tolive{and} of 2.4 \AA, 2.1 \AA\ and 2.0 \AA\ for Fe-S, P-P, and P-S, respectively.}
A full BFGS relaxation is then performed by imposing hydrostatic pressure (by introducing a target stress tensor for the desired hydrostatic pressure \ttred{in {\sc CASTEP}}) 
arriving at a relaxed structure and corresponding energy, and enthalpy.} 
About 1500 initial structures were generated across the potential energy surface (PES) at each pressure point. 
The method cannot ensure that the global minimum is found as the search cannot \tteal{be} exhaustive \tolive{within a} finite computation time, 
as is well known for global minima searches \cite{Pickard2011AbSearching, Kennedy2002_SWARM,wang2012calypso,wu2013adaptive_genetic}. 
Yet, the adequacy of our search is supported by \tteal{us} finding \tolive{the} expected \tteal{stable phases} several times, the so-called 'marker' structures. 
In our case, they are the known low-pressure (LP) \tolive{phase}, and the first and second high-pressure phases (HP-I and HP-II), as characterised in the literature \cite{2018_Haines_highPressure_XRD_FePS3}.


\section{Results and Discussion}

\begin{figure*}[htbp]
\centering
\includegraphics[width=1.0\textwidth,angle=0]{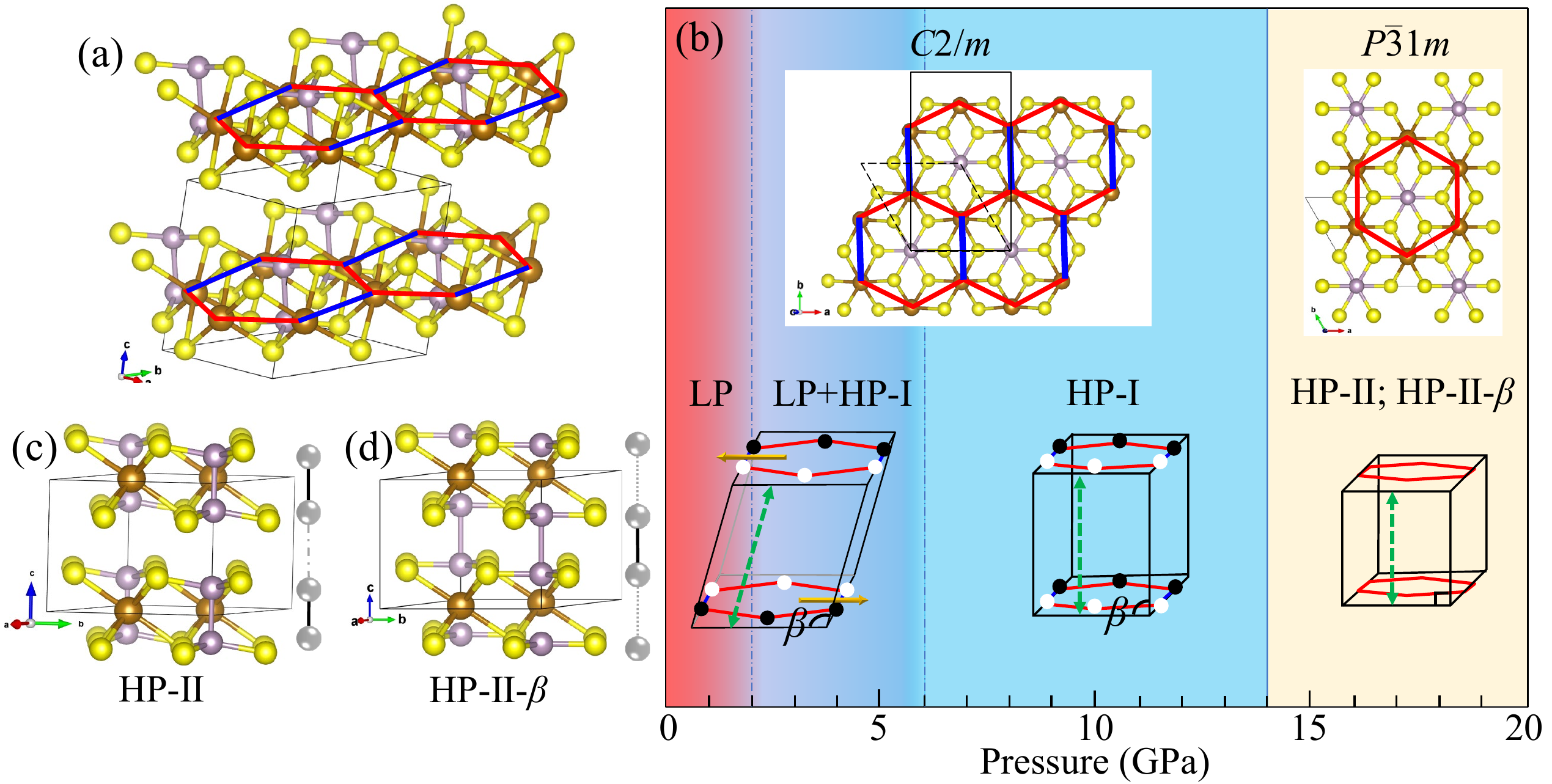}
\caption{\tred{Crystal structure of various FePS$_3$ phases and phase diagram 
vs pressure. (a), (c) and (d) show the LP, HP-II and HP-II-$\beta$ structures.}
  (b) Pressure phase diagram Ref.\cite{2021_Coak_MagPhase_FePS3} and random structure 
search results in this \tteal{paper}.
\tblue{The red and blue rods indicate the ferromagnetic (FM) and antiferromagnetic (AFM) coupling between the nearest neighbouring Fe ions, respectively. 
The black and white dots represent the magnetic moment on Fe$^{2+}$ pointing up or down, normal to the $ab$ plane. 
The LP and HP-I both host zigzag FM chains along $a$ axis and the neighbouring chains within each plane are AFM coupled. 
From LP to HP-I, the interlayer AFM coupling changes into FM instead. 
Below when we refer to them as (LP, AFM) and (HP-I, AFM) in contrast with the non-spin-polarized phases.
The transition pressure value is sensitive to the magnitude of Hubbard $U$, for example it is $\sim 7.4$ GPa for $U=2.50$ eV (see Fig. \ref{PhaseTran}), and $\sim 11$ GPa for $U=3.75$ eV figure (see Appendix~\ref{app:approx}).}
}
\label{FePS3_all_structure}
\end{figure*}

With the AIRSS search we manage to reproduce the known phases and in addition predict a new one \tteal{in} the high-pressure region. 
The novel structure shares the same \tolive{template} of the honeycomb and space group symmetry as HP-II,
\tolive{though it distinguishes} itself with the \tolive{characteristic} P$_2$ bonding.
The P atoms form stronger chemical bonds between the neighbouring layers instead of the intralayer \tteal{configuration} of previously proposed structures.
\tteal{There are hints of a similar evolution in the related material NiPS$_3$ \cite{2021_Ma_NiPS3}.}
\tolive{The now fully connected 3D structure therefore differs from the weakly bound two-dimensional layers found in the HP-II phase.}
We refer to this phase as HP-II-$\beta$ \tteal{going forward}. 
Metastable phases with diverse building blocks are also predicted but in this work we focus on the phases in close proximity to the \tred{minimal enthalpy line, 
which happen to be close to experimental findings except for the mentioned HP-II-$\beta$ phase.}
Fig.\ref{FePS3_all_structure} summarizes the LP, HP-II, HP-I and HP-II-$\beta$ phases and a detailed discussion of the evolution with pressure is given in the subsequent section.

\subsection{Pressure-induced Crystalline Phase Transitions}

To establish the context \tolive{in which} the crystalline structure evolves with pressure, 
we start with the LP phase. 
FePS$_3$ \tteal{has been subjected} to a full X-ray structural characterisation since 1973, being the first among the family\cite{klingen1973_FePX3_structure}. 
The neighbouring layers stack together and the bulk crystallized in the monoclinic space group $C2/m$ with an ideal $\beta$ angle, compared with the Mn and Ni counterparts \cite{Ouvrard1985_TMPS3_structures}, which suggest the least distortion within each layer. 
At ambient pressure, FePS$_3$ is a Mott insulator with \tolive{a} band gap of $\sim$ 1.5 eV. 
Across the family, the complicate\tolive{d} competition among exchanges and anisotropy leads to different types of antiferromagnetic order with different directions for the collinear axes of the spin moments (Ising versus Heisenberg magnetic Hamiltonian). 
FePS$_3$ at ambient pressure has Ising-type AFM order with zigzag ferromagnetic (FM) chains along the $a$ axis coupled antiferromagnetically along the $b$ axis. 
The moments are normal to the $ab$ planes \cite{klingen1973_FePX3_structure}. 
Being a 2D AFM with correlated physics, FePS$_3$ has been studied extensively \tgreen{in recent} years with pressure being the tuning parameter. 

Fig. \ref{FePS3_all_structure} (a) displays the crystalline structure of LP FePS$_3$ with the Fe$^{2+}$ honeycomb being illustrated by a rigid rod. 
In the inset of Fig. \ref{FePS3_all_structure} (b), the obtained spin up and down moments are indicated via black and white dots, respectively, \tred{defining ferromagnetically (FM) coupled lines of Fe atoms (red rods), which are antiferromagnetically ordered \tteal{in respect to} each other (blue rods). 
This is a striking result for a Fe essentially honeycomb arrangement, 
which being bipartite would not \tgreen{be expected to induce} the kind of frustrations \tteal{which give} rise to the observed arrangement. 
It is, however, a known and expected result from the LP phase related to frustrations due to first vs longer range effective exchange couplings \cite{2020_Wildes_Biquadratic_FePS3}.
}
  
The magnetic configuration breaks the $C_3$ rotational symmetry and leads to slightly elongated Fe hexagons in the LP phase. 
The blue rods corresponding to \tolive{the} AFM coupling are 3.448 \AA\ in length, 
while the red ones representing FM coupling are 3.426 \AA. 
The intersite exchange within and in-between the planes is mediated through the surrounding P$_2$S$_6$ clusters, with P atoms \tolive{centered within} the distorted Fe hexagons.
\tred{The shear in the interlayer coupling can also be considered to be behind the C3 symmetry breaking, but it is observed (see below) the \tolive{symmetry} breaking remains 
when the shear is removed by pressure.}
Defining the formula unit (f.u.) as [FePS$_3$], the primitive cell contains 2 f.u. in the cell, as indicated by the dotted line in the upper-left inset of Fig.~\ref{FePS3_all_structure} (b). 
The conventional cell \tolive{consistent} with previous literature involves 4 f.u. of [FePS$_3$], shown in solid black lines. 
For consistency with previous works, we adopt the conventional description of the lattice parameters with $a$ = 5.947 \AA, $b$ = 10.300 \AA, 
$c$ = 6.722 \AA\ and $\beta$ = 107.2$^{\circ}$ for the LP. 
  \tred{The symmetry breaking distorts the Fe hexagons, showing 
sides along the zigzag spin chains that are \tblue{shorter} than the
ones perpendicular \tolive{to the chain}, by a ratio of \tblue{0.65 \% at ambient pressure}.
  The $b/a$ ratio is, however, only very slightly affected,  
staying very close to the ideal $b/a\ = \sqrt{3}$.}

With increas\tolive{ing} pressure, the crystalline structure of LP evolves into new phases but some features of the LP are \tteal{retained}. 
The HP-I phase share\tolive{s} the same crystalline building block as LP, but differs in the stacking angle $\beta$ as neighbouring layers in LP slide towards each other.
The $\beta$ angle turns nearly 90$^{\circ}$ in HP-I. 
In addition, though the in-plane magnetic configuration has been preserved, 
the interlayer coupling has changed from AFM to FM \cite{2021_Coak_MagPhase_FePS3}. 
Both phases crystallise in the space group of $C2/m$. 
Experimentally a coexistence region \tolive{of both LP and HP-I} is observed from 2 to 6 GPa, \tblue{from the 
powder diffraction data without pressure medium
\cite{2018_Haines_highPressure_XRD_FePS3,2021_Coak_MagPhase_FePS3}. 
Nevertheless, no coexistence was seen in the single crystal, 
or in the powder data with He pressure medium \cite{2023_Jarvis_comparative_single_powder_FePS3}.}
We discuss the transition from LP to HP-I in Section~\ref{LP_HP-I_sliding}.

\begin{figure}[htbp]
    \centering
    \includegraphics[width=0.75\textwidth, angle=0]{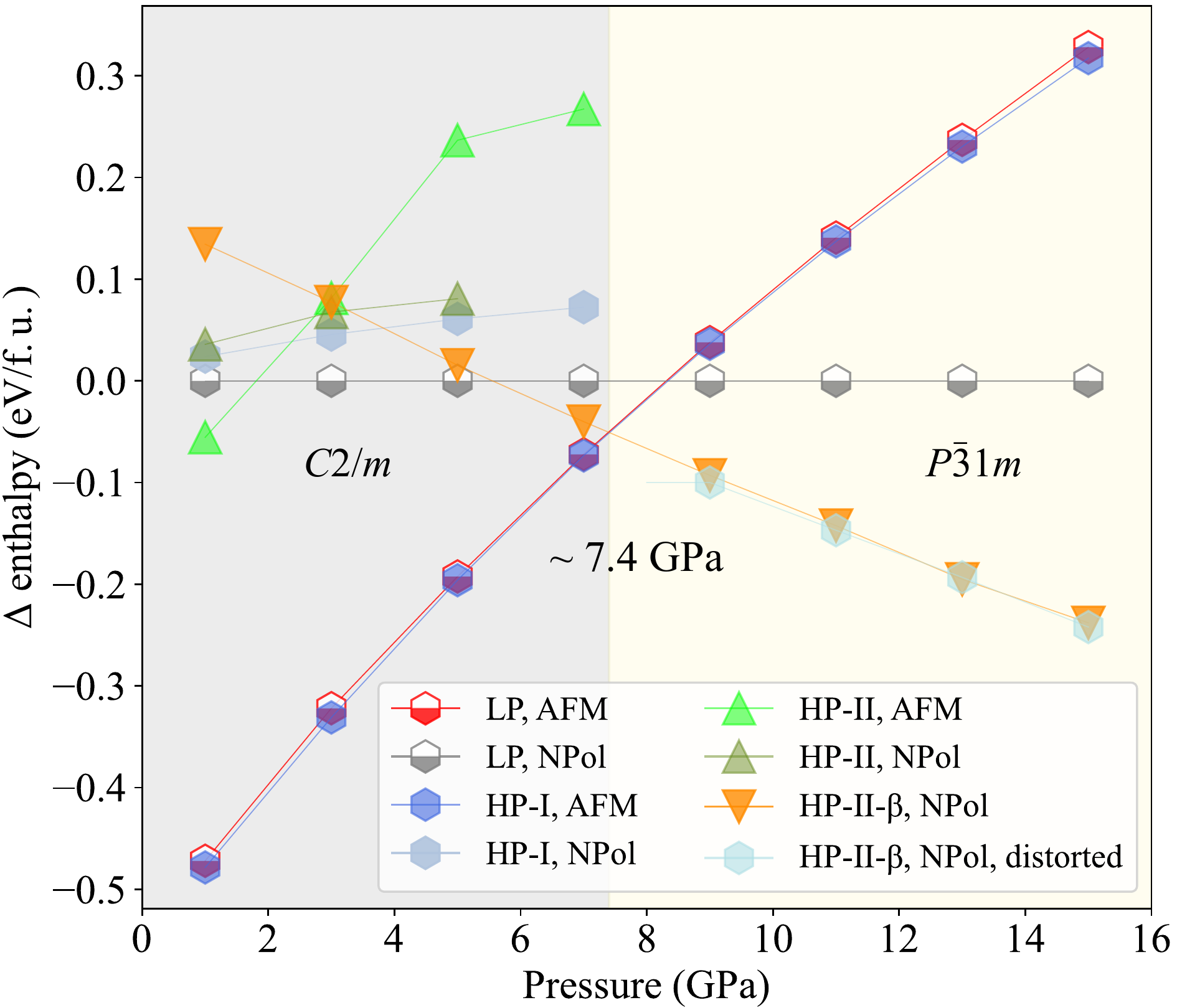}
    \caption{Pressure dependence of the enthalpy difference $\Delta H = H - H_{\mathrm{ref}}$ of various phases with respect to the reference non-spin-polarized low-pressure phase (LP, NPol, half-grey hexagons). 
    $\Delta H$ is shown for: the low-pressure phase with antiferromagnetically coupled spin chains (LP, AFM, half-red hexagons), 
    high-pressure phase I, both AFM (blue hexagons) and non-polarised (HP-I, NPol, full light blue hexagons) and, high-pressure phase II, HP-II, both in the N\'eel AFM state (light green triangles) and NPol (dark green triangles) and the non-polarised HP-II-$\beta$ phase (orange triangles). 
    The figure shows that below a transition pressure of $\sim 7.4$ GPa the antiferromagnetic LP and HP-I phases are very close in enthalpy, 
    LP being very slightly favored at low pressures, but very close to unshearing to the HP-I phase (see Fig.~\ref{enthalpy_beta}). 
    Above the transition pressure the HP-II phase becomes the stable one in its $\beta$ form, that is, with interlayer P-P bonds. 
    \ttred{The structure was relaxed allowing for a distortion breaking the triangular symmetry. The relaxation recovered the symmetry up to a residual distortion of 0.006 \AA\ in the difference of inequivalent Fe-Fe nearest neighbour distances.}}
\label{PhaseTran}
\end{figure}

At the higher pressure region, the long-range magnetic order is suppressed and the Fe$^{2+}$ cations \tolive{form} perfect hexagons within the plane, as indicated in the upper-right inset of Fig.~\ref{FePS3_all_structure} (b). 
The recovery of the $C_3$ rotational symmetry at the centre of \tolive{the} Fe-hexagons leads to the symmetry crossover transition from $C2/m$ to $P\overline{3}1m$. 
\hl{The previously proposed HP-$\rm{II}$ and the predicted 
HP-$\rm{II}$-$\beta$ share the same \tolive{intralayer S-Fe-S template}.} 
The side-view \hl{for these two phases} is shown in 
Fig.~\ref{FePS3_all_structure} (c) \hl{and (d), respectively}.
The P$_2$ bonding behaviour alters the dimensionality in HP-II from that in HP-II-$\beta$. 
It offers an ideal platform to explore the effect of dimensionality on the electronic and magnetic properties, which will be discussed in detail \tolive{in Section~\ref{level2:electronic}}.

In order to understand how these phases evolve with pressure, we performed first-principles geometry optimisation at various pressure points. 
The enthalpy-pressure phase diagram is summarised in Fig.~\ref{PhaseTran}. 
The enthalpy for the non-polarised LP phase has been chosen as the reference line. 
The \tblue{spin polarisation} is shown to be essential in \tolive{establishing} the relative stability of the various phases, 
with a very significant energy scale of up to half an eV/f.u. stabilizing the LP and HP-I phases.
However, given their full polarization, both phases compete very closely in enthalpy, with differences LP and HP-I now \tolive{on} the meV/f.u. scale.


\subsubsection{Coexistence of the LP and HP-I}
\label{LP_HP-I_sliding}

\begin{figure}[htbp]
\centering
    \includegraphics[width=0.7\textwidth,angle=0]{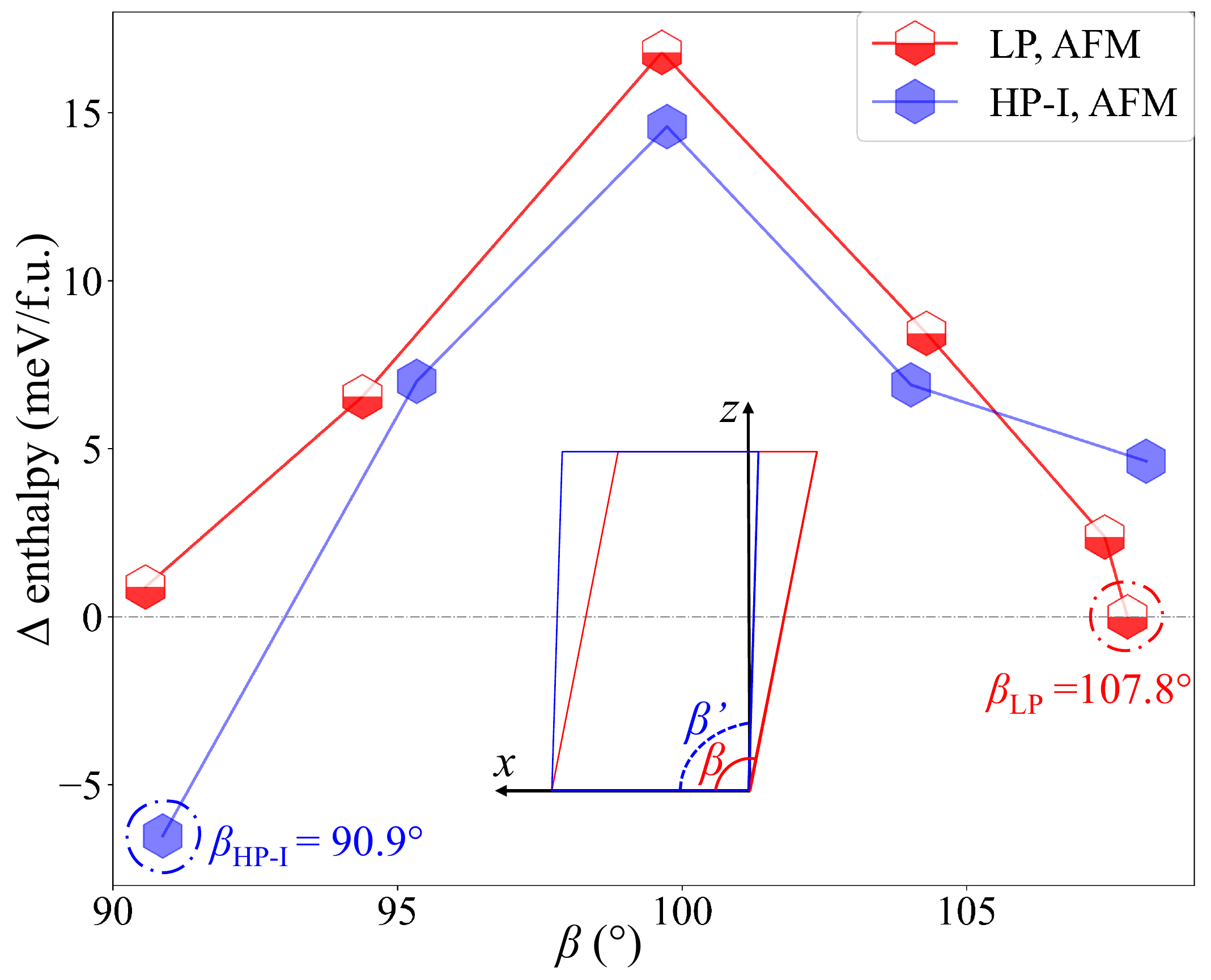}
    \caption{Enthalpy difference as a function of the $\beta$ angle for the LP and HP-I AFM phases of FePS$_3$ at 1.0 GPa. 
    The enthalpy of LP relaxed at a fixed $\beta_{LP}$ = 107.8$^\circ$ is set as reference.
    The difference between both curves at the same pressure and shear is the different interlayer spin arrangement of the LP and HP-I phases.}
\label{enthalpy_beta}
\end{figure}

The transition from LP to HP-I occurs via the \tolive{relative} sliding of \tolive{the}
neighbouring layers and the change of interlayer magnetic coupling from AFM to FM. 
  We display the enthalpy evolution in response to the interlayer sliding in LP and HP-I
as a function of the $\beta$ angle in Fig.~\ref{enthalpy_beta} at 1 GPa of pressure.
  The inset in the figure illustrates the shear in terms of the lattice viewed 
from the $b$ axis. 
  For each phase, we manually slide the neighbouring layers and then relax the 
geometry with a fixed $\beta$ angle.
  The enthalpy of the fully optimised LP phase at $\beta$ = 107.8 $^\circ$ \tolive{is}
set as \tolive{the} reference.
  \tred{The first conclusion extracted from this plot is that already at this pressure
the almost unsheared phase would be the stable one.}
  Howewer, a significant energy barrier ($\sim$ 15 meV per f.u.) becomes very apparent 
in the figure for both phases, which could explain the experimentally observed 
coexistence of LP and HP-I shown in Fig.~\ref{FePS3_all_structure} (b) between 
2 and 6 GPa, due to kinetic effects in the relaxation of the loading. 
  In addition, experimentally the pressure cannot be ideally hydrostatic and 
\tred{phase coexistence might persist over a pressure range, depending on experimental conditions 
\tolive{as it seems to have happened in experiments on powder samples\cite{2018_Haines_highPressure_XRD_FePS3}.}}


\subsubsection{The Dimensionality Change in HP-II and HP-II-$\beta$}
\label{HP-II_HP-II_dimensionality}

\begin{figure}[htbp]
\centering
    \includegraphics[width=0.65\textwidth,angle=0]{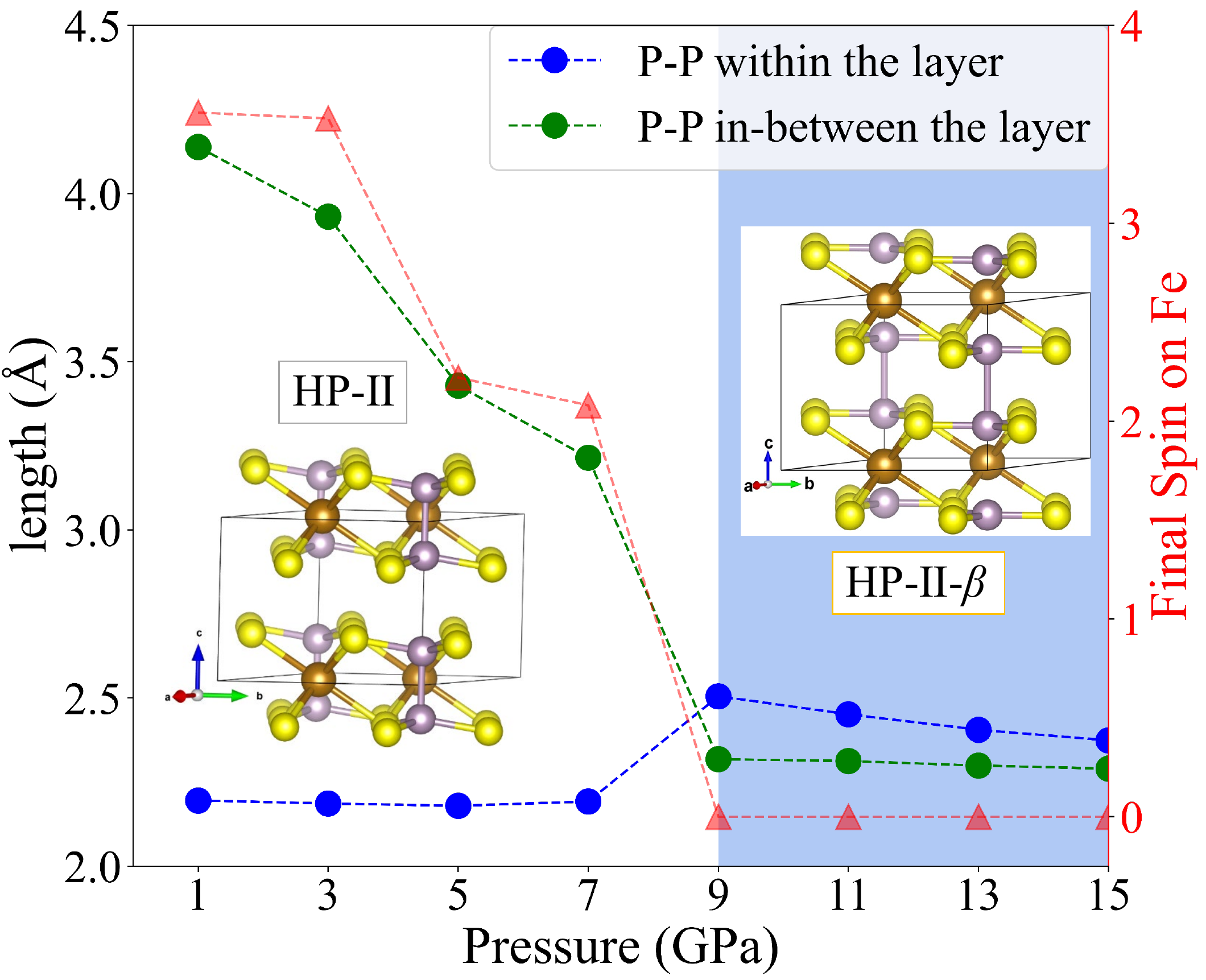}
    \caption{Evolution with the pressure of P-P interatomic distances for P nearest neighbors along the direction perpendicular to the layers, in the $P\overline{3}1m$ FePS$_3$ phase. 
    Intralayer (interlayer) distance is shown as blue (green) circles. 
    Red triangles show the localised spin on a Fe atom.}
\label{HP2_lattice}
\end{figure}

Back in Fig.~\ref{PhaseTran}, there is a symmetry transition at $P \sim$ 7.4 GPa from $C2/m$ (LP, HP-I) to $P\overline{3}1m$, which is qualitatively similar to the previous experimental findings. 
Such symmetry change also occurs when the bulk is reduced to monolayer thickness \cite{2020_Neal_TMPS3_symmCrossover_Layers}. 
The trigonal symmetry for every single layer at that pressure is recovered by releasing the constraint of the monoclinic $\beta$ angle. 
\ttred{When starting from a distorted geometry (HP-I), the structure relaxes towards the trigonal phase. 
The average in-equivalent Fe-Fe nearest neighbour distance is about 3.350 \AA\ and the difference is 0.006 \AA\ for the residual distortion. 
The enthalpy for the HP-II-$\beta$ phase with residual distortion is also shown in Fig.~\ref{PhaseTran}.}

\ttred{For the two trigonal phases,} 
HP-II and HP-II-$\beta$ distinguish from one another in the P-P intra- and interlayer bonding behaviours. 
In order to investigate whether they are degenerate in energy or relax into the same local minimum after geometry optimisation, we explore the evolution of crystal structure including lattice parameters \tolive{at the applied pressure}, 
especially following the interatomic distance of P-P within and in-between the layers. 

Fig.~\ref{HP2_lattice} shows the evolution of P-P distances within the layer (blue circles) and between the layers (green) for the relaxed HP-II phase from 1 to 15 GPa in steps of 2 GPa. 
The prediction is that HP-II transforms into the HP-II-$\beta$ phase \tred{around 9 GPa}. 
The novel HP-II-$\beta$ phase is more 3D-like, compared \tolive{to the HP-II phase}. 
\tred{As far as we understand, 
\tolive{it is difficult to experimentally determine}
the detailed P positions. 
In addition, further experimental evidence may be impeded by the substantial kinetic energy barrier expected for the change in connectivity, which implies P-P chemical bond restructuring.}

Fig.~\ref{HP2_lattice} was obtained with the PBE functional, corrected by the vdW Trachenko-Scheffler approach, and a value of the Hubbard $U$ = 2.5 eV on the Fe $d$ orbitals \ttred{including spin polarization for the Fe atoms}. 
We extensively explore \tred{the sensitivity of such a transition to the simulation parameters and approximations \tolive{mentioned} in Appendix~\ref{app:approx}}. 
The key prediction is that the dimensionality change\tolive{s} from 2D to 3D under pressure. 
Yet, increased values of Hubbard $U$ postpone the transition to higher pressure, 
\tred{bringing it into closer agreement with experiments, and thereby suggesting} 
that the correlation strength in the FePS$_3$ system is quite substantial and makes a difference.

\subsubsection{The Dynamical Stability of HP-II and HP-II-$\beta$}
\label{HP-II_dynamical}

\begin{figure}[htbp]
\centering
    \includegraphics[width=0.8\textwidth,angle=0]{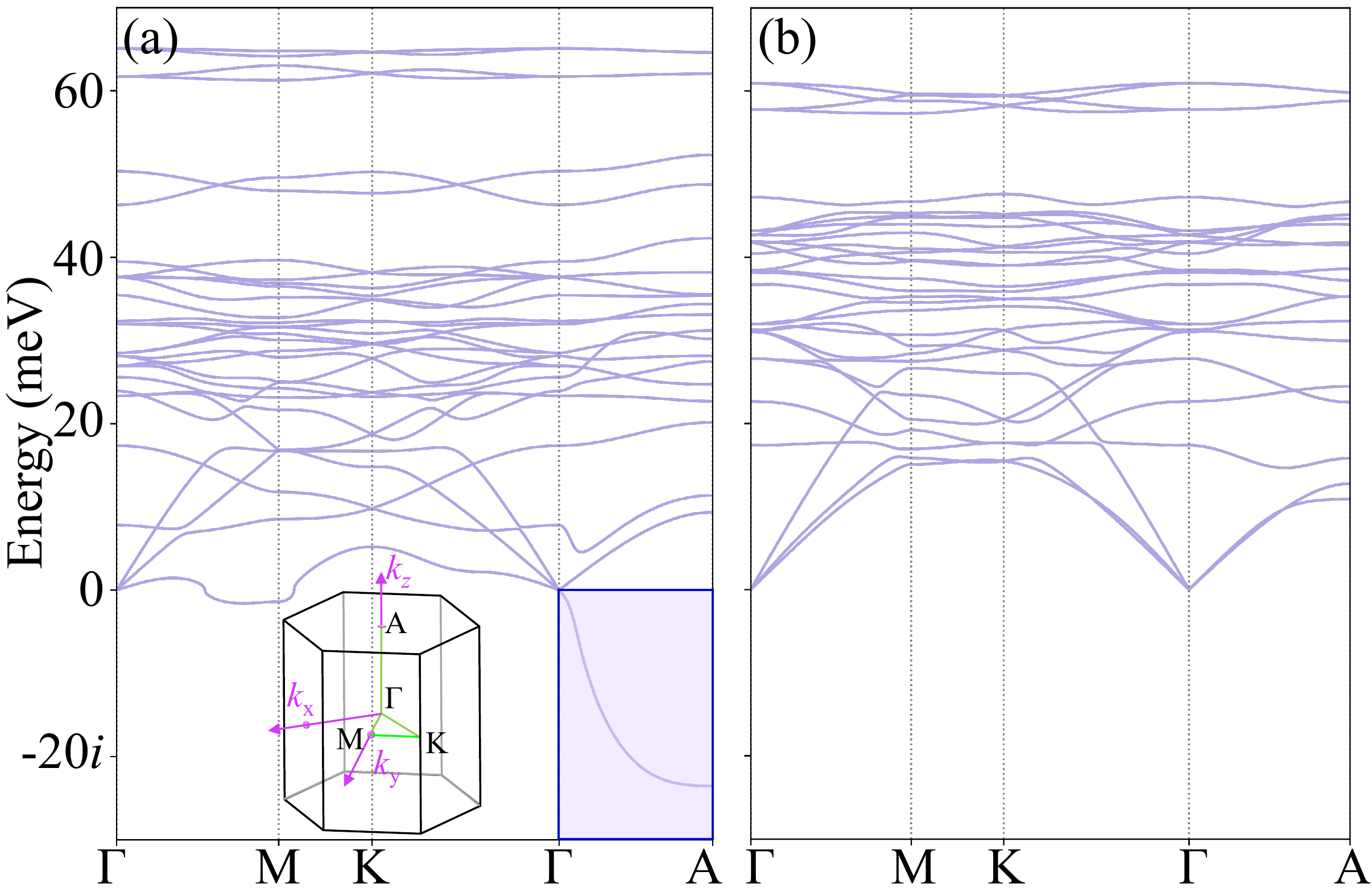}
    \caption{The phonon spectrum calculated within DFT+$U$ (Fe $d$: $U=2.5$ eV) for fully relaxed (a) HP-II and (b) HP-II-$\beta$ phases at 7.0 GPa. The selected high symmetry path has been illustrated as an inset in (a).}
    \label{phonon_HP2_HP2-beta}
\end{figure}


\ttred{We have also explored the dynamical stability for HP-II and HP-II-$\beta$ phases under the same pressure before the transition occurs, here 7.0 GPa.
\tviolet{The phonon spectrum has been calculated 
using the finite displacement method \cite{PhysRevLett.78.4063} in conjunction with nondiagonal supercells \cite{PhysRevB.92.184301}.}
Multiple commensurate supercells have been first constructed
\tviolet{where atoms are perturbed from equilibrium positions}
And then SCF calculations have been performed to evaluate the force-constant matrix.
The finite displacement method allows the use of
\tviolet{
any electronic structure theory to obtain the lattice dynamics of FePS$_3$, taking the vdW interaction and Fe $d$-orbital correlation into account}. 
A \tviolet{$3 \times 3 \times 3$ q-point grid} has been adopted \tviolet{to sample the dynamical matrix} of HP-II and HP-II-$\beta$ \tviolet{phases}.}

\ttred{
The phonon dispersions for the two phases are shown in Fig. \ref{phonon_HP2_HP2-beta} along a selected high-symmetry path within the Brillouin zone.
Though there is no imaginary phonon at the zone center ($\Gamma$ point) for HP-II, 
the imaginary part along $\Gamma$ to A, \tviolet{highlighted in the shaded region,} suggests certain instability for the HP-II structure along the \tviolet{vertical direction, which means that the HP-II phase tends to distort itself along the direction to lower its energy}. 
The symmetry constraint of the $P\overline{3}1m$ space group might give the reason why HP-II can still be found via the structure search and geometry relaxation before the transition occurs. 
\tviolet{By contrast, there is no imaginary phonon frequency across the whole Brillouin zone for the HP-II-$\beta$ phase}, 
suggesting that this predicted new phase featuring P-P interlayer bonding geometry is indeed dynamically stable. 
This is consistent with the fact that \tviolet{a more compact structure along the vertical direction will be favored at high pressure. }
\tviolet{In addition, it is also worth noting that} the volume of HP-II is about 8.8 \% larger than that of HP-II-$\beta$, mainly arising from the $c$ lattice parameter being 5.407 \AA\ for the former while being 4.915 \AA\ for the latter, more 3D-connected phase. 
The P-P bonding length within the layer versus that in-between the layer is 2.192/3.215 for HP-II, and 2.605/2.310 for HP-II-$\beta$.
\tviolet{The significant volume shrinkage along the vertical direction and the rearrangement of P-P bonding lead to the fact that} the \tviolet{high-frequency} optical modes in HP-II are \tviolet{softened in HP-II-$\beta$.} 
\tviolet{We have observed that} the optical modes around 63 meV refer to \tviolet{strong} vibration\tviolet{s} of P and S atoms with all Fe atoms remaining static.
Those around 47 meV refer to the vibration\tviolet{s} of predominantly S atoms. 
From \tviolet{the} 2D \tviolet{layered structure} (HP-II) to \tviolet{the} 3D bonded structure (HP-II-$\beta$), the vibration modes in FePS$_3$ within the $ab$ plane have similar energy while those related to the perpendicular direction \tviolet{are} heavily affected.
}


\subsection{Electronic Properties of FePS$_3$}
\label{level2:electronic}

\subsubsection{LP and HP-I}

\begin{figure}[htbp]
\centering
\includegraphics[width=0.65\textwidth,angle=0]{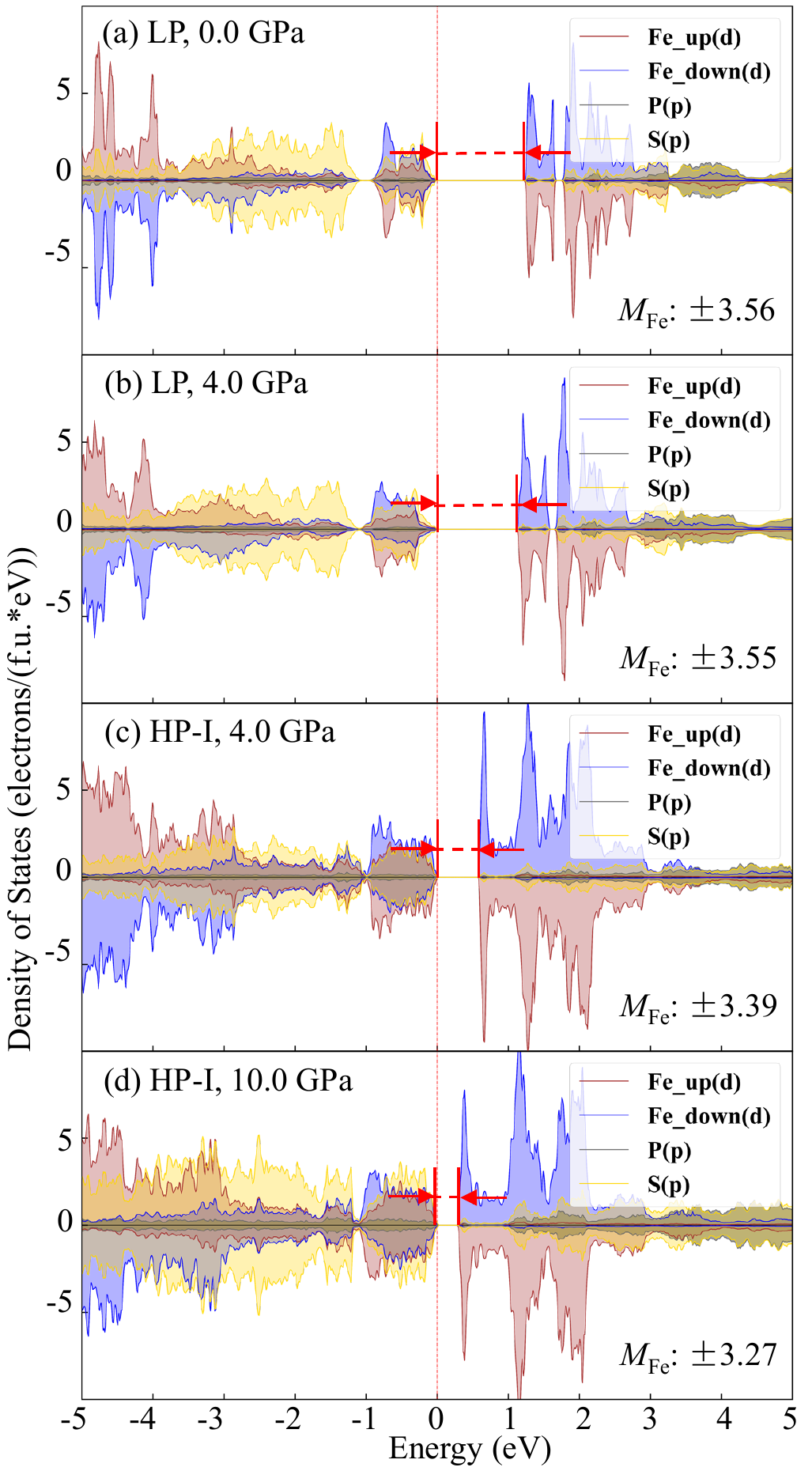}
\caption{Projected density of states (PDOS) as calculated within DFT+$U$ 
(Fe $d$: $U=2.5$ eV), post-processed with the OptaDOS package \cite{2012_nicholls_optados,2014_morris_optados}, showing
the projection of Kohn-Sham states onto the Fe $3d$ states for spin-up (positive sign, red) and down (negative sign, blue). 
  Grey PDOS indicate P 3$p$ and yellow is for S $3p$.
  Panels (a) and (b) for LP at 0.0 GPa and 4.0 GPa (b), respectively,
(c) and (d) show HP-I at 4.0 GPa and 10.0 GPa.
\tblue{$M_{\rm{Fe}}$ refers to the final spin moments on Fe.}
}
\label{pdos_insulating}
\end{figure}

  We have also investigated the evolution of electronic properties 
along with the structural transition in response to external pressure. 
  The projected density of states (PDOS) are calculated with DFT+$U$ 
(Fe $d$: $U=2.5$ eV) methodology, using CASTEP code and then post-processed 
with the OptaDOS package \cite{2012_nicholls_optados,2014_morris_optados}. 
  In the LP and HP-I phases, the spin-up and spin-down magnetic arrangements
for all Fe atoms are equivalent as they both stabilize with long-range AFM order. 
  Fig.~\ref{pdos_insulating} displays the PDOS involving the projection of 
Kohn-Sham states onto the spin-up and \tolive{spin-}down Fe 3$d$ orbitals, together with 
the ones for P 3$p$ and S $3p$, 
for LP at 0.0 GPa and 4.0 GPa (b), 
and for HP-I at 4.0 GPa \tolive{(c)} and 10.0 GPa (d). 
  \tred{The apparent perfect spin-up-down compensation is characteristic 
of net zero moments in AFM arrangements. 
  It should be \tteal{noted}, however, that, if projected on separate Fe
atoms, they show mostly up (or down) according to their net polarization 
obtained and displayed in Fig.~\ref{FePS3_all_structure}.}

  \tred{The obtained electronic structure is consistent with the attributed
Mott insulating state, with a clear split between occupied and unoccupied
Fe $3d$ orbitals, the weight of the S anions staying clear of the bands
around the Fermi level, thereby ruling out a charge-transfer character \tteal{for}
the insulating state. 
  This fact is especially clear for LP, but less clean cut for the HP-I phase,
and in both cases, P states significantly mix with the Fe $3d$ states, 
complicating the modelling.
  This latter fact is probably what is behind the less idealized 
parameterizations needed in spin models \tblue{\cite{2012_Wildes_magnon_SpinExchange_FePS3}}, which 
give rise to the frustration effects \tteal{\tolive{that} result in} the observed spin arrangements, 
instead of the pure N\'eel state
expected from the 
bipartite honeycomb Fe substructure.
  The obtained electronic structure for this phase is also consistent 
with the fact that the energy scales for the definition of spin moments, 
the high spin state for Fe$^{2+}$ ($S$ = 2) and the effective spin-spin 
interactions dominate over 
any crystal field splitting parameter, $\varDelta$, that would tend to reduce 
the net spin on individual Fe atoms.}

The obtained band gap, as determined by the bottom of the conduction and the top of the valence Kohn-Sham bands, for the LP phase at 0.0 GPa is about 1.2 eV. 
This is qualitatively consistent with \tblue{previous optical measurements of \tolive{the} band gap \cite{1986_Grasso_optical_TMPS3}.} 
  \tred{It is well known that band gaps are not quantitatively predicted 
from this level of theory.
  Although the DFT$+U$ correction (including sensible, phenomenological 
values of $U$) somewhat improves their reliability, we will follow them 
here qualitatively as support for consistency and \tgreen{an} indicator of trends.}

  With increas\tolive{ing} pressure, the band gap for LP gradually shrinks down to 
$\sim 1.1$ eV at 4.0 GPa. 
  However, when LP transforms into HP-I, the band gap becomes 
significantly smaller, $\sim$ 0.56 eV for the HP-I at 4.0 GPa. 
  The HP-I phase remains insulating at \tolive{the applied pressure $P$} = 10.0 GPa with a 
reduced band gap of $\sim 0.27$ eV. 
  Though the increased pressure would increase the crystal field splitting 
energy $\varDelta$ and thus affect the competition between $\varDelta$ and the 
exchange energy $J$, the high-spin state of Fe$^{2+}$ ($S$ = 2) is still 
maintained in both the LP and HP-I phases within the pressure range 
being discussed.

\subsubsection{HP-II and HP-II-$\beta$}

\begin{figure}[htbp]
\centering
\includegraphics[width=0.65\textwidth,angle=0]{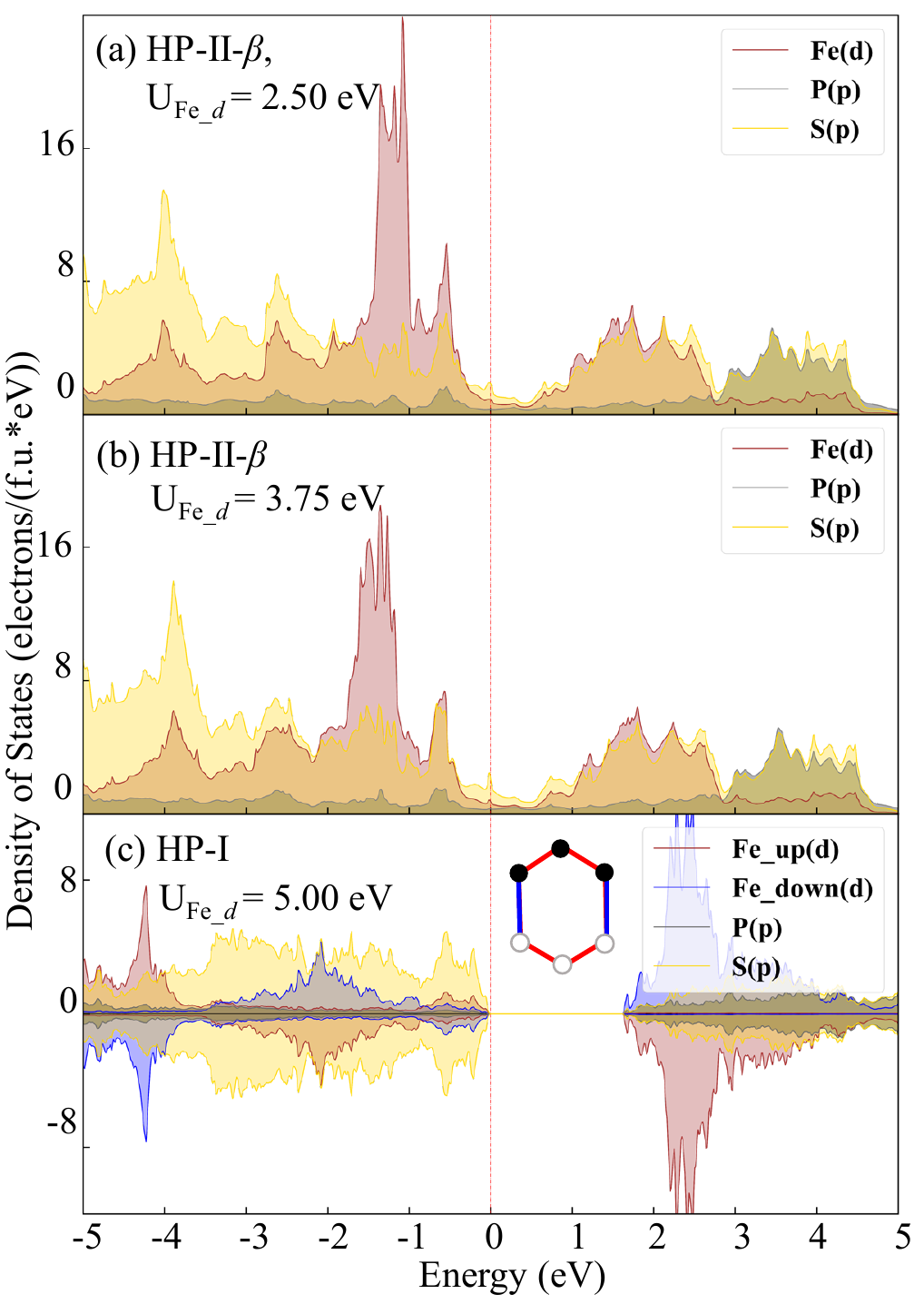}
\caption{PDOS for HP-II-$\beta$ [panels (a) and (b)] and HP-I (c), 
respectively at $P=11.0$ GPa (same conventions as for Fig.~\ref{pdos_insulating}).
Hubbard $U=2.50$ eV (a), 3.75 eV (b), and 5.00 eV (c) on the Fe $3d$ orbitals.}
\label{pdos_11GPa_U}
\end{figure}

\tred{Figure~\ref{pdos_11GPa_U} shows the same PDOS decomposition as Fig.~\ref{pdos_insulating} now for $P=11.0$ GPa and the HP-II phases.
Panel (a) shows the HP-II-$\beta$ phase, which appears clearly beyond the insulator to metal transition, \tolive{using} the same level of theory as employed so far.
No spin decomposition is obtained.}
\tred{The dimensionality difference, introduced by the distinctive interlayer P bonding behaviour, affects the electronic structure across the Fermi level. 
With continuous contributions from P and S states, the metallicity also significantly affecting the distribution of the 3$d$ orbitals of Fe.}

\tred{There is uncertainty in the most suitable \tolive{choice for the} value \tolive{of} the Hubbard $U$.
The experimental IMT observed at higher pressures (see Fig.~\ref{FePS3_all_structure}) seems to indicate stronger 
correlation effects than for other Fe compounds.}
In Fig.~\ref{pdos_11GPa_U} we explore the effect of on-site Coulomb repulsion by tuning the Hubbard $U$ on Fe $d$ orbitals from 2.5 to 3.75 and 5.0 eV. 
The geometries are relaxed at 11.0 GPa with different $U$, and the corresponding PDOS are then calculated as shown.
\tred{The increased $U$ values push the insulator to metal transition to higher pressures, in closer agreement with experiments, the insulating phase always remaining HP-I (panel c) with the same spin arrangements (inset).}
The increased electron-electron repulsion seems to inhibit 
the formation of strong chemical inter-layer P-P bonds, while retaining the long-range order (LRO) zigzag structures within planes. 
\tred{Interestingly, whenever the gap closes, we obtain the $\beta$ phase.}

\begin{figure}[htbp]
\centering
\includegraphics[width=0.68\textwidth,angle=0]
{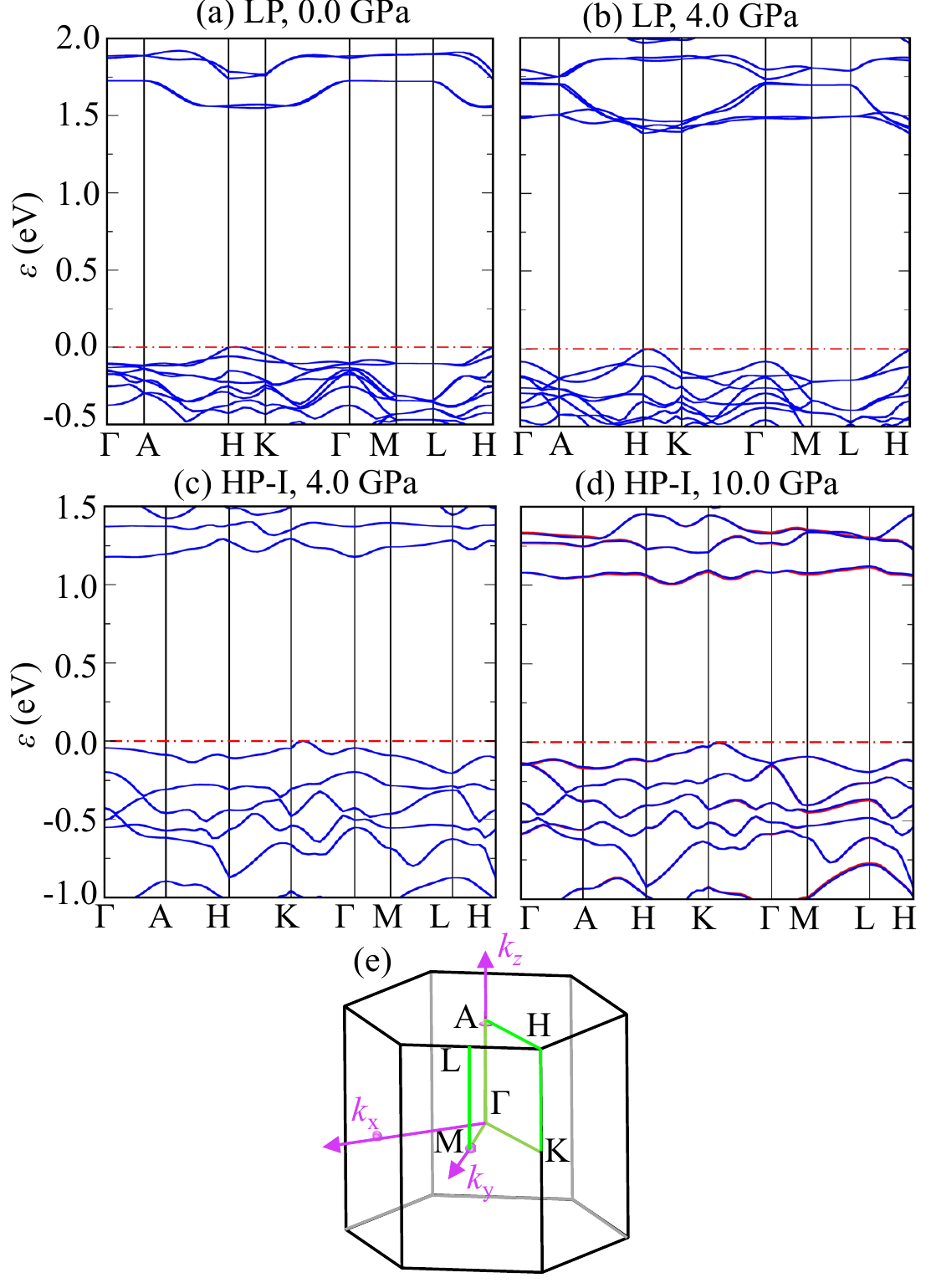}
\caption{Band structures for (a) LP at 0 GPa and (b) 4.0 GPa, and for (c) HP-I at 4.0 GPa and (d) 10.0 GPa, 
\tblue{and (e) a Brillouin zone diagram indicating the symmetry points defining the band dispersion plots.}}
\label{bands_all}
\end{figure}

\tred{The fact that the calculations lose spin polarization upon metalization has to be properly interpreted.
  The DFT$+U$ calculations rely on \tolive{the} mean-field treatments of the electronic problem 
(even at the Hubbard level) and cannot describe short-range spin correlations.}
  Coak $et\ al.$\cite{2021_Coak_MagPhase_FePS3} observe from their experiments
that short-range order (SRO) magnetism persists in the high-pressure region,
resembling that of the HP-I magnetic configuration.
It is an interesting finding, compatible with the physics close to the Mott transition. 
\tred{There could be ways to query the calculations in that direction, by introducing constrained spins by hand at selected locations, and seeing the spin texture decaying from those sites in large-simulation-box settings.
  In an AFM situation the procedure is however quite arbitrary, since the individual magnetic moments develop for the electrons associated \tteal{with} a given atom,
which is always an ill-posed proposition.
  Higher levels of theory, such as dynamical-mean-field theory (DMFT) \cite{2006_kotliar_electronic_DMFT} 
would also help in this regard, but they represent a substantially increased 
computational and theoretical effort, well beyond the scope of this work.}

\begin{figure}[htbp]
\centering
    \includegraphics[width=0.7\textwidth,angle=0]{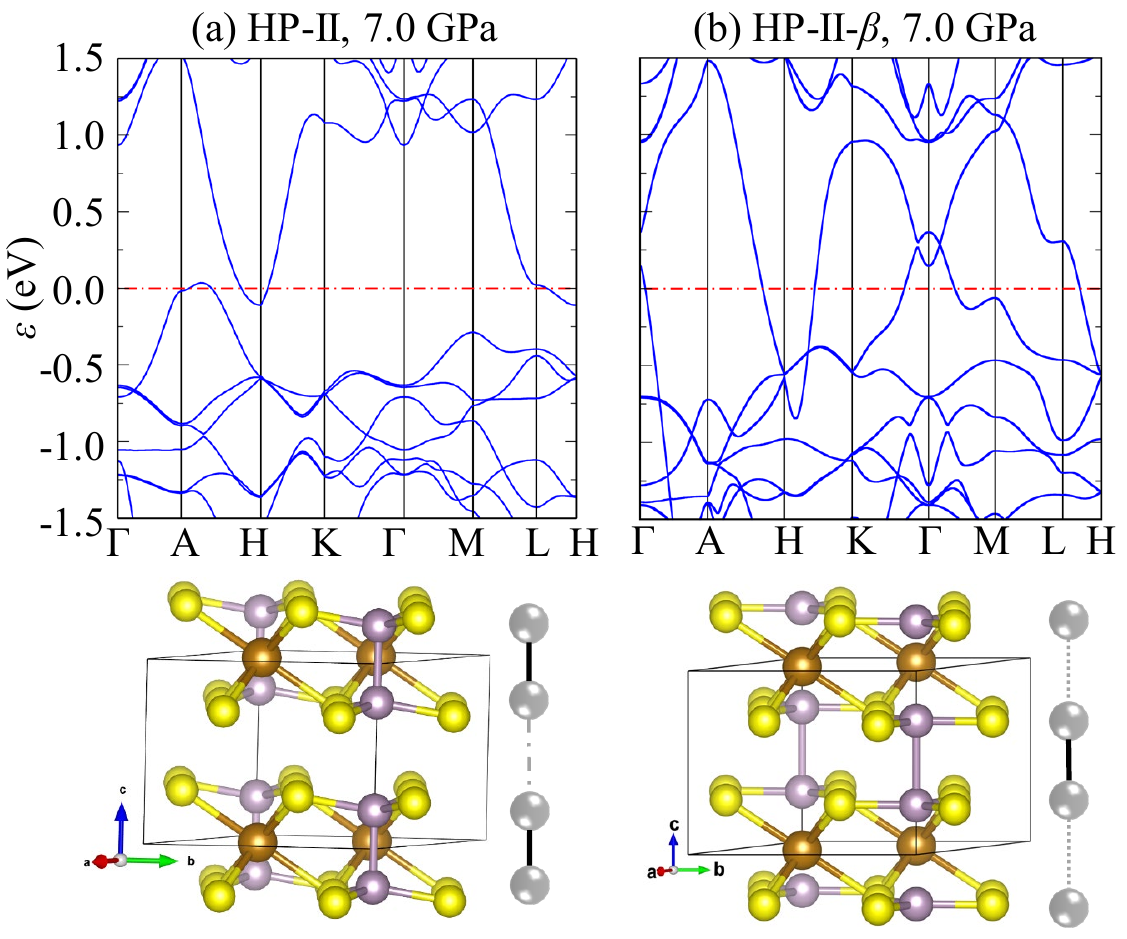}
    \caption{Band structures for (a) HP-II and (b) HP-II-$\beta$ phases at 7.0 GPa (both $P\overline{3}1m$), close to the transition point, with structural sketches under each corresponding panel.}
    \label{bands_HP2_HP2-beta}
\end{figure}

In addition to the displayed PDOS for various phases, Figs.~\ref{bands_all} and \ref{bands_HP2_HP2-beta} show the corresponding band-structure dispersion relations along paths crossing selected high-symmetry points.
Fig.~\ref{bands_all} shows \tolive{these} for LP and HP-I, from 0.0 GPa to 10.0 GPa.
The $\Gamma$-A and M-L Brillouin-zone segments are along the $k_z$ direction, 
(nearly) perpendicular to the layer planes in real space. 
The bands along \tolive{these} directions are quite flat, as expected for weakly interacting 2D layers.

\tred{Fig.~\ref{bands_HP2_HP2-beta} shows the band dispersion for the metallic high-pressure phases, very close to the insulator-to-metal transition ($P=7.4$ GPa, as obtained for $U+2.5$ eV), in both its originally proposed layered form HP-II and its 3D HP-II-$\beta$ version.
The metallization is much clearer in the latter, the former
representing a weak semimetal.}
In Fig. \ref{bands_HP2_HP2-beta} (a), electron and hole pockets emerge near the Fermi level \tblue{along the high symmetry path H-A}. 
It suggests that the HP-II phase would transit through a semimetal phase before becoming fully metallic at slightly higher pressures. 
\tred{From our results, however, it is difficult to ascertain whether there could be such an intermediate region of stability.}

Several bands cross the Fermi level in HP-II-$\beta$, 
as can be seen in Fig.~\ref{bands_HP2_HP2-beta} (b). 
%
Our results suggest that the metallicity would emerge at the high-pressure region regardless of the dimensionality collapse in the $c$ axis. 
Our calculations offer another scenario in this case. 
\ttred{It should be noted that for the high-pressure phase reported in \cite{2022_Raman_DFT_FePS3,2022_Zeng_FePS3_MagTran_DFT} the band structure is actually very similar to the one we obtain for HP-II-$\beta$, in spite of their not reporting on the dimensionality change implied by the $\beta$ phase.
} 
\tred{The $c$\tolive{-axis} collapse and P-P interlayer bonds appearing in the HP-II-$\beta$ phase should affect electronic transport significantly.} 
Further investigation is needed to understand the effect of dimensionality crossover on \tred{electronic transport.
\tcyan{It would require the deployment of }more sophisticated techniques (like DMFT \cite{2006_kotliar_electronic_DMFT}, whenever accessible to these system sizes) \tred{to describe} the effect of the \tteal{strong} electron correlations induced by the strength of on-site Coulomb repulsion for 3$d$ orbitals}.


\section{Conclusions}

With AIRSS random structure search, we \tcyan{have reproduced }the previously proposed LP, HP-I and HP-II \tcyan{phases,} and a novel HP-II-$\beta$ phase. 
LP and HP-I crystallise in the monoclinic space group $C2/m$ while HP-II and HP-II-$\beta$ stabilize in the trigonal one $P\overline{3}1m$. 
The HP-II-$\beta$ is intrinsically different from HP-II in that the P atoms form stronger chemical bonds in-between the neighbouring layers. 
The full simulation of the enthalpy-pressure phase diagram, taking into consideration \tblue{spin polarisation}, is found to be consistent with the previous experimental observations.

We rationalize the coexistence region of the LP and HP-I phases by quantifying the energy barriers for neighbouring planes to slide against each other, regardless of the interplanar coupling being antiferromagnetic (LP) or ferromagnetic (HP-I). 
\tcyan{Despite the gradual decrease of the band gap size, both phases remain insulating in this pressure region.}

At higher pressure, the symmetry crossover from $C2/m$ to $P\overline{3}1m$ \tcyan{takes place concurrent with \ttred{the} emergence of} metallicity. 
The intra- or inter-layer P-P bonding affects the energy bands near the Fermi level, and \tcyan{might be responsible for a possibly different explanation for} the origin of metallicity. 
\ttred{
The predicted dynamically stable HP-II-$\beta$ phase defines a scenario for FePS$_3$ turning metallic while becoming 3D-connected under high pressure.}

The magnetic moments on the transition metal sites are suppressed when the system turns from quasi-2D to the 3D limit. 
\hl{The strength of the on-site Coulomb repulsion also influences the dimensionality change. 
The actual correlation strength deserves further investigation as it could facilitate quantitative theory in the low-dimensional $TM\rm{P}X_3$ and other AFM Mott systems.}

Considering the complexity of correlation effects in the system and the competition among exchange and anisotropy, more effort is needed to further our understanding of the nature of the insulator-to-metal transition and how dimensionality is involved in the process.
This work has been carried out with the DFT+U \tcyan{technique, which is insensitive to} environmental factors \tcyan{like nature of} pressure medium. 
The theoretical modelling can be expected to guide future experimental explorations in the $TX\rm{P}X_3$ compound family and could possibly be extended to other 2D magnets, where pressure or other tuning parameters could tune the phases to be novel metal or unconventional superconductors.

\section*{Acknowledgement}

The authors would like to thank C.J. Pickard, M.J. Coak, 
D.M. Jarvis, C.R.S. Haines, H. Hamidov, C. Liu, X. Zhang and A.R. Wildes for their generous help and discussions.   
This work was carried out with the support of the Cambridge Service for Data-Driven Discovery (CSD3) and the UK Materials and Molecular Modelling Hub (MMM Hub).

\paragraph{Funding information}

S.S.S. and S.D. acknowledge support from Department for Business, Energy and Industrial Strategy, UK (BEIS Grant Number: G115693) and Cavendish Laboratory.
S.D. acknowledges a scholarship to pursue doctoral research from the Cambridge Trust China Scholarship Council. 
S.C. acknowledges financial support from the Cambridge Trust and from the Winton Programme for the Physics of Sustainability.
B.M. acknowledges financial support from a UKRI Future Leaders Fellowship (Grant No. MR/V023926/1), from the Gianna Angelopoulos Programme for Science, Technology, and Innovation, and from the Winton Programme for the Physics of Sustainability.  
E. A. acknowledges funding from Spanish MICIN through grant PID2019-107338RB- C61/AEI/10.13039/501100011033, 
as well as a Mar\'{\i}a de Maeztu award to Nanogune, Grant CEX2020-001038-M funded by MCIN/AEI/ 10.13039/501100011033.




\begin{appendix}

\section{Convergence tests}
\label{app:tests}

\begin{figure}[htbp]
    \centering
    \includegraphics[width=1.0\textwidth,angle=0]{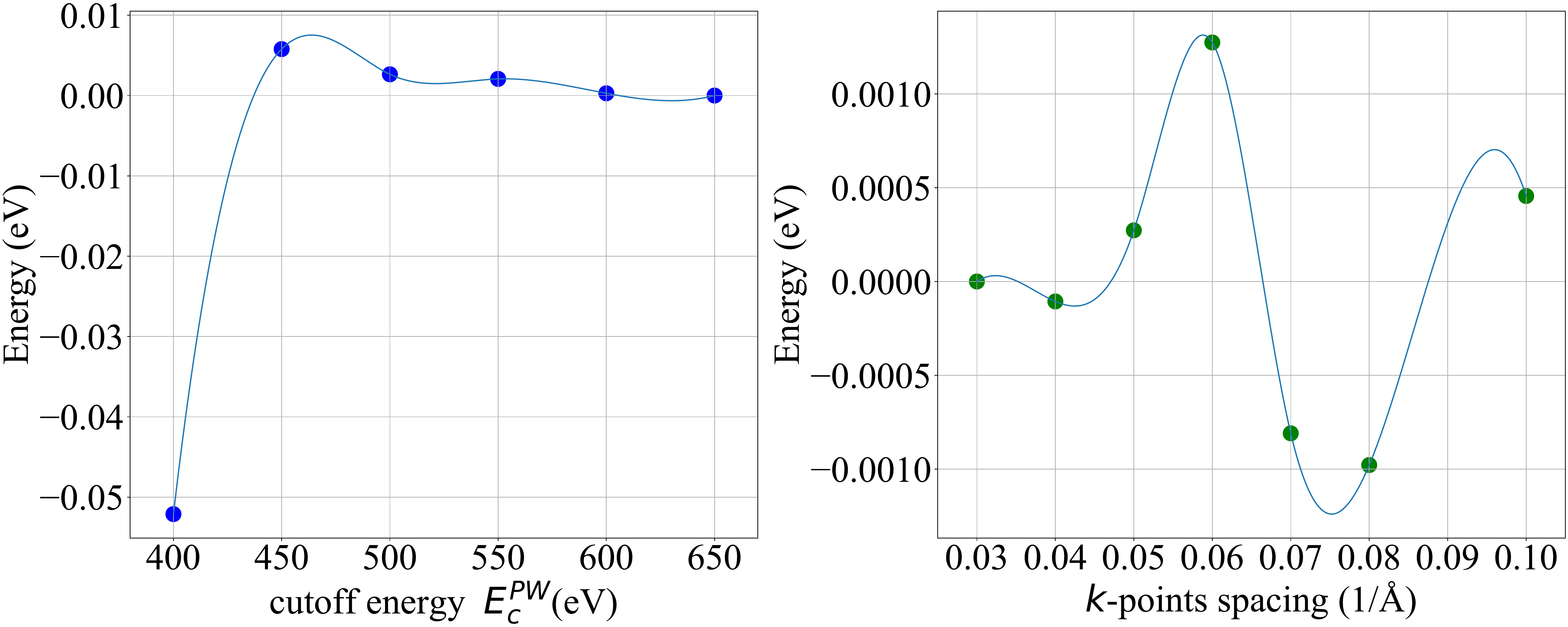}
    \caption{Convergence with plane-wave energy cutoff $E_c^{PW}$ of the 
    total energy of the LP phase of FePS$_3$ at ambient pressure (left), and with \textit{\textbf{k}}-point spacing (right). 
    The vertical axis of energy has been rescaled so that the energies calculated with the highest energy cutoff  ($E_c^{PW}$ = 650 eV) and the highest \textit{\textbf{k}}-point density (spacing = 0.03 1/\AA\ ) corresponds to 0 eV.
    }
    \label{convergence}
\end{figure}

We have performed the convergence tests for the plane-wave cutoff energy ($E_c^{PW}$) and \textit{\textbf{k}}-points sampling within the framework of static single-point calculations using CASTEP. 
The LP phase of FePS$_3$ at ambient pressure has been chosen to perform the tests. 
The total energy dispersions as a function of $E_c^{PW}$ and the maximum spacing between each \textit{\textbf{k}}-point are displayed in Fig.~~\ref{convergence}.
Based on the convergence tests, the $E_c^{PW}$ = 550 eV and k-points sampling of 0.03 \AA$^{-1}$ along each axis has been set for the rest of all calculations, 
including the high-pressure phases.


\section{Sensitivity of dimensional crossover at high pressure for various approximations}
\label{app:approx}

\begin{figure*}[htbp]
\includegraphics[width=1.0\textwidth,angle=0]{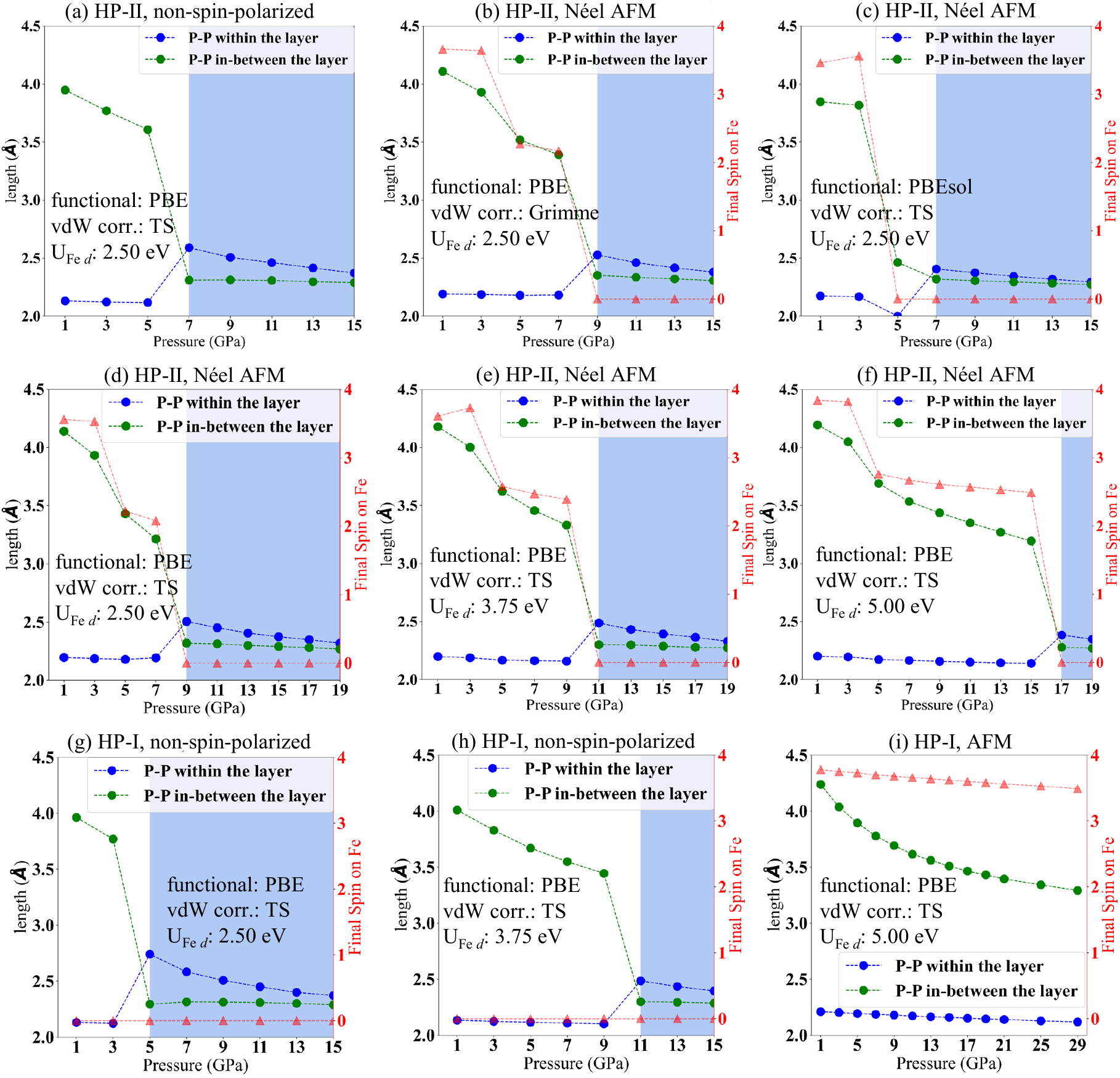}
	\caption{Evolution of P-P interatomic distances within (blue) and in-between (green) the layers in response to simulated hydrostatic pressure for the HP-II (a-f) and HP-I (g-i) phase under different simulation scenarios.}
	\label{HP2_lattice_supp}
\end{figure*}

We also investigated the sensitivity of pressure-induced P-P from adjacent layers forming stronger bonds with respect to the choice of simulation parameters in detail.
Considering the fact that FePS$_3$ is a layered vdW compound with Fe being able to host finite magnetic moments, 
we postulate a few starting scenarios in HP-II phase and then fully relax the structure with CASTEP. 
The P-P interatomic distances within each double sulfur layer and in between the neighbouring layers are evaluated from the post-optimized phases. 
Fig. \ref{HP2_lattice_supp} summarizes the two
P-P interatomic distances as a function of pressure, 
with the former in blue and the latter in green.

In Fig. \ref{HP2_lattice_supp} (a), the non-spin-polarized HP-II phase has been relaxed from 1 to 15 GPa with a step size of 2 GPa.
The functional is chosen as PBE\cite{Perdew1996GeneralizedSimple}, 
and the vdW correction methodology follows the Tkatchenko-Scheffler (TS)'s approach \cite{2009_TS_correction}. 
There are the parameters that we have discussed and utilized for most of the calculations in the main text.
We also compared the S. Grimme's semiempirical approach (2006) \cite{2006_grimme_semiempirical} (\ttred{equivalent to} the DFT-D2 in the Vienna $ab\ initio$ simulation package (VASP) \cite{2008_VASP_DFT}) to tackle the vdW interactions.
It can be seem from Fig. \ref{HP2_lattice_supp} (b) that the P-P interlayer dimerization with pressure will not be affected by the choice of vdW correction methodology qualitatively. 
Similarly, the PBESol\cite{2008_perdew_PBESol} as functional has been explored in Fig. \ref{HP2_lattice_supp} (c), without qualitative effect on the transition.

\tblue{
In Fig. \ref{HP2_lattice_supp} (d-f), 
we explored the effect of Hubbard $U$ on Fe $d$ orbitals 
at 2.5, 3.75 and 5.0 eV. 
The strength of Hubbard $U$ has the most prominent influence. 
The transition point having been postponed to as high as around 17 GPa when $U$ is 5.0 eV.
In addition, we also explored the effect of symmetry in Fig. \ref{HP2_lattice_supp} (g-i).
We create the supercell with 4 f.u. inside one simulation cell 
from HP-II phase 
to allow for zigzag FM chains along $a$ axis 
being antiferromagnetically coupled to each other within each layer. 
The relaxed structures by symmetry analysis 
shall be classified as HP-I phases instead. 
The formation of interlayer P-P dimer is more sensitive to the correlation effect when the $C3$ rotation symmetry has been broken. 
The neighbouring layers remain to be gapped via vdW interaction.
}

To conclude, P atoms tend to form shorter and stronger bonds across neighbouring layers at higher pressure. 
The high-pressure phase stabilizes in HP-II-$\beta$ energetically within the DFT+U methodology. 
The choice of different vdW corrections to the final energy and density functionals barely affects the system to stabilize in the HP-II-$\beta$ at high-pressure region.
Meanwhile, the Hubbard $U$ on Fe $d$ orbitals influences the formation of interlayer P-P bonds substantially, which is consistent with the picture of correlated electrons in FePS$_3$.


\end{appendix}




\bibliography{ref}

\begin{thebibliography}{10}
\providecommand{\url}[1]{\texttt{#1}}
\providecommand{\urlprefix}{URL }
\expandafter\ifx\csname urlstyle\endcsname\relax
  \providecommand{\doi}[1]{doi:\discretionary{}{}{}#1}\else
  \providecommand{\doi}{doi:\discretionary{}{}{}\begingroup
  \urlstyle{rm}\Url}\fi
\providecommand{\eprint}[2][]{\url{#2}}

\bibitem{Gong2019_2DMagnet_Review}
C.~Gong and X.~Zhang,
\newblock \emph{Two-dimensional magnetic crystals and emergent heterostructure
  devices},
\newblock Science \textbf{363}, eaav4450 (2019),
\newblock \doi{10.1126/science.aav4450}.

\bibitem{2023_FePX3_water_splitting_application}
S.~Sharma, H.~M. Zeeshan, M.~Panahi, Y.~Jin, M.~Yan, Y.~Jin, K.~Li, P.~Zeller,
  A.~Efimenko, A.~Makarova, D.~Smirnov, B.~Paulus \emph{et~al.},
\newblock \emph{{Stability of van der Waals FePX$_3$ materials (X: S, Se) for
  water-splitting applications}},
\newblock 2D Materials \textbf{10}(1), 014008 (2022),
\newblock \doi{10.1088/2053-1583/ac9c15}.

\bibitem{1966_Mermin_Wagner_Theorem}
N.~D. Mermin and H.~Wagner,
\newblock \emph{Absence of ferromagnetism or antiferromagnetism in one- or
  two-dimensional isotropic heisenberg models},
\newblock Phys. Rev. Lett. \textbf{17}, 1133 (1966),
\newblock \doi{10.1103/PhysRevLett.17.1133}.

\bibitem{1967_Hobenberg}
P.~C. Hohenberg,
\newblock \emph{Existence of long-range order in one and two dimensions},
\newblock Phys. Rev. \textbf{158}, 383 (1967),
\newblock \doi{10.1103/PhysRev.158.383}.

\bibitem{2016_Fe3GeTe2}
H.~L. Zhuang, P.~R.~C. Kent and R.~G. Hennig,
\newblock \emph{Strong anisotropy and magnetostriction in the two-dimensional
  stoner ferromagnet fe$_3$gete$_2$},
\newblock Phys. Rev. B \textbf{93}, 134407 (2016),
\newblock \doi{10.1103/PhysRevB.93.134407}.

\bibitem{Gong2017_Cr2Ge2Te6}
C.~Gong, L.~Li, Z.~Li, H.~Ji, A.~Stern, Y.~Xia, T.~Cao, W.~Bao, C.~Wang,
  Y.~Wang, Z.~Q. Qiu, R.~J. Cava \emph{et~al.},
\newblock \emph{{Discovery of intrinsic ferromagnetism in two-dimensional van
  der Waals crystals}},
\newblock Nature \textbf{546}(7657), 265 (2017),
\newblock \doi{10.1038/nature22060},
\newblock \eprint{1703.05753}.

\bibitem{LeFlem1982_TMPS3_OldTheoModel}
G.~{Le Flem}, R.~Brec, G.~Ouvard, A.~Louisy and P.~Segransan,
\newblock \emph{{Magnetic interactions in the layer compounds MPX$_3$ (M = Mn,
  Fe, Ni; X = S, Se)}},
\newblock J. Phys. Chem. Solids \textbf{43}(5), 455 (1982),
\newblock \doi{10.1016/0022-3697(82)90156-1}.

\bibitem{2006_Wildes_MnPS3}
A.~R. Wildes, H.~M. R{\o}nnow, B.~Roessli, M.~J. Harris and K.~W. Godfrey,
\newblock \emph{{Static and dynamic critical properties of the
  quasi-two-dimensional antiferromagnet MnPS$_3$}},
\newblock Phys. Rev. B \textbf{74}(9), 094422 (2006),
\newblock \doi{10.1103/PhysRevB.74.094422}.

\bibitem{2016_Lee_IsingTypeOrder_FePS3}
J.~Lee, S.~Lee, J.~H. Ryoo, S.~Kang, T.~Y. Kim, P.~Kim, C.~H. Park, J.~G. Park
  and H.~Cheong,
\newblock \emph{{Ising-Type Magnetic Ordering in Atomically Thin FePS$_3$}},
\newblock Nano Letters \textbf{16}(12), 7433 (2016),
\newblock \doi{10.1021/acs.nanolett.6b03052}.

\bibitem{Halperin2019_HMW_theorem}
B.~I. Halperin,
\newblock \emph{On the hohenberg--mermin--wagner theorem and its limitations},
\newblock J. Stat. Phys. \textbf{175}(3), 521 (2019),
\newblock \doi{10.1007/s10955-018-2202-y}.

\bibitem{martin2016thin}
L.~W. Martin and A.~M. Rappe,
\newblock \emph{Thin-film ferroelectric materials and their applications},
\newblock Nat. Rev. Mater. \textbf{2}(2), 1 (2016),
\newblock \doi{10.1038/natrevmats.2016.87}.

\bibitem{2022_ProximityEffect_Monolayer_MPX3}
K.~Zollner and J.~Fabian,
\newblock \emph{{Proximity effects in graphene on monolayers of
  transition-metal phosphorus trichalcogenides $M\mathrm{P}{X}_{3}$
  $(M:\mathrm{Mn}, \mathrm{Fe}, \mathrm{Ni}, \mathrm{Co}, \mathrm{and} X:
  \mathrm{S}, \mathrm{Se})$}},
\newblock Phys. Rev. B \textbf{106}, 035137 (2022),
\newblock \doi{10.1103/PhysRevB.106.035137}.

\bibitem{2022_Zhou_SpinShearCoupling_vdWAFM}
F.~Zhou, K.~Hwangbo, Q.~Zhang, C.~Wang, L.~Shen, J.~Zhang, Q.~Jiang, A.~Zong,
  Y.~Su, M.~Zajac, Y.~Ahn, D.~A. Walko \emph{et~al.},
\newblock \emph{Dynamical criticality of spin-shear coupling in van der waals
  antiferromagnets},
\newblock Nat. Commun. \textbf{13}, 6598 (2022),
\newblock \doi{10.1038/s41467-022-34376-5}.

\bibitem{1986_First_Cuprate}
J.~G. Bednorz and K.~A. M{\"{u}}ller,
\newblock \emph{{Possible high $T_c$ superconductivity in the Ba-La-Cu-O
  system}},
\newblock Zeitschrift f{\"{u}}r Phys. B Condens. Matter \textbf{64}(2), 189
  (1986),
\newblock \doi{10.1007/BF01303701}.

\bibitem{1988_Cuprate_120K}
Z.~Z. Sheng and A.~M. Hermann,
\newblock \emph{{Bulk superconductivity at 120 K in the Tl-Ca/Ba-Cu-O system}},
\newblock Nature \textbf{332}(6160), 138 (1988),
\newblock \doi{10.1038/332138a0}.

\bibitem{2014_kumar_pressure_SC_LaFeAsO}
R.~S. Kumar, J.~J. Hamlin, M.~B. Maple, Y.~Zhang, C.~Chen, J.~Baker, A.~L.
  Cornelius, Y.~Zhao, Y.~Xiao, S.~Sinogeikin \emph{et~al.},
\newblock \emph{{Pressure-induced superconductivity in LaFeAsO: The role of
  anionic height and magnetic ordering}},
\newblock Appl. Phys. Lett. \textbf{105}(25), 251902 (2014),
\newblock \doi{10.1063/1.4904954}.

\bibitem{2009_Imai_FeSe_SC_pressure}
T.~Imai, K.~Ahilan, F.~Ning, T.~M. McQueen and R.~J. Cava,
\newblock \emph{{Why does undoped FeSe become a high-$T_c$ superconductor under
  pressure?}},
\newblock Phys. Rev. Lett. \textbf{102}(17), 177005 (2009),
\newblock \doi{10.1103/PhysRevLett.102.177005}.

\bibitem{2021_PNAS_pressure_highT_SC_FeSe}
L.~Deng, T.~Bontke, R.~Dahal, Y.~Xie, B.~Gao, X.~Li, K.~Yin, M.~Gooch,
  D.~Rolston, T.~Chen, Z.~Wu, Y.~Ma \emph{et~al.},
\newblock \emph{{Pressure-induced high-temperature superconductivity retained
  without pressure in FeSe single crystals}},
\newblock Proc. Natl. Acad.Sci. \textbf{118}, 2021 (2021),
\newblock \doi{10.1073/pnas.2108938118}.

\bibitem{2021_NatComm_pressure_SC_MnSe}
T.~L. Hung, C.~H. Huang, L.~Z. Deng, M.~N. Ou, Y.~Y. Chen, M.~K. Wu, S.~Y.
  Huyan, C.~W. Chu, P.~J. Chen and T.~K. Lee,
\newblock \emph{{Pressure induced superconductivity in MnSe}},
\newblock Nat. Commun. \textbf{12}, 5436 (2021),
\newblock \doi{10.1038/s41467-021-25721-1}.

\bibitem{2014_NatComm_SC_near_AFM_CrAs}
W.~Wu, J.~Cheng, K.~Matsubayashi, P.~Kong, F.~Lin, C.~Jin, N.~Wang, Y.~Uwatoko
  and J.~Luo,
\newblock \emph{{Superconductivity in the vicinity of antiferromagnetic order
  in CrAs}},
\newblock Nat. Commun. \textbf{5}, 5508 (2014),
\newblock \doi{10.1038/ncomms6508}.

\bibitem{2022_Pressure_induced_SC_AuTe2Br}
E.~Cheng, X.~Shi, L.~Yan, T.~Huang, F.~Liu, W.~Ma, Z.~Wang, S.~Jia, J.~Sun,
  W.~Zhao, W.~Yang, Y.~Xu \emph{et~al.},
\newblock \emph{{Critical topology and pressure-induced superconductivity in
  the van der Waals compound AuTe$_2$Br}},
\newblock npj Quantum Mater. \textbf{7}, 93 (2022),
\newblock \doi{10.1038/s41535-022-00499-7}.

\bibitem{2001_Monthoux_SC_2D_3D}
P.~Monthoux and G.~G. Lonzarich,
\newblock \emph{Magnetically mediated superconductivity in quasi-two and three
  dimensions},
\newblock Phys. Rev. B \textbf{63}, 054529 (2001),
\newblock \doi{10.1103/PhysRevB.63.054529}.

\bibitem{2020_NatPhys_SC_LowD}
B.~Sacépé, M.~Feigel’man and T.~M. Klapwijk,
\newblock \emph{Quantum breakdown of superconductivity in low-dimensional
  materials},
\newblock Nature Physics \textbf{16}, 734 (2020),
\newblock \doi{10.1038/s41567-020-0905-x}.

\bibitem{2016_2DMagnet_outlook}
J.~G. Park,
\newblock \emph{{Opportunities and challenges of 2D magnetic van der Waals
  materials: magnetic graphene?}},
\newblock J. Condens. Matter Phys. \textbf{28}(30), 301001 (2016),
\newblock \doi{10.1088/0953-8984/28/30/301001}.

\bibitem{2018_Burch_Magnetism_vdW_Materials}
K.~S. Burch and J.~G. Mandrus, D.and~Park,
\newblock \emph{{Magnetism in two-dimensional van der Waals materials}},
\newblock Nature \textbf{563}(7729), 47 (2018),
\newblock \doi{10.1038/s41586-018-0631-z}.

\bibitem{2016Wang_MnPX3_spinCrossover}
Y.~Wang, Z.~Zhou, T.~Wen, Y.~Zhou, N.~Li, F.~Han, Y.~Xiao, P.~Chow, J.~Sun,
  M.~Pravica, A.~L. Cornelius, W.~Yang \emph{et~al.},
\newblock \emph{{Pressure-Driven Cooperative Spin-Crossover, Large-Volume
  Collapse, and Semiconductor-to-Metal Transition in Manganese(II) Honeycomb
  Lattices}},
\newblock J. Am. Chem. Soc. \textbf{138}(48), 15751 (2016),
\newblock \doi{10.1021/jacs.6b10225}.

\bibitem{2019_MnPX3_NiPX3_pressure_IMT}
H.~S. Kim, K.~Haule and D.~Vanderbilt,
\newblock \emph{{Mott Metal-Insulator Transitions in Pressurized Layered
  Trichalcogenides}},
\newblock Phys. Rev. Lett. \textbf{123}, 236401 (2019),
\newblock \doi{10.1103/PhysRevLett.123.236401}.

\bibitem{2018_Haines_highPressure_XRD_FePS3}
C.~Haines, M.~Coak, A.~Wildes, G.~Lampronti, C.~Liu, P.~Nahai-Williamson,
  H.~Hamidov, D.~Daisenberger and S.~Saxena,
\newblock \emph{{Pressure-Induced Electronic and Structural Phase Evolution in
  the van der Waals Compound FePS$_3$}},
\newblock Phys. Rev. Lett. \textbf{121}(26), 266801 (2018),
\newblock \doi{10.1103/PhysRevLett.121.266801}.

\bibitem{2018_Wang_pressure_SC_FePX3}
Y.~Wang, J.~Ying, Z.~Zhou, J.~Sun, T.~Wen, Y.~Zhou, N.~Li, Q.~Zhang, F.~Han,
  Y.~Xiao, P.~Chow, W.~Yang \emph{et~al.},
\newblock \emph{{Emergent superconductivity in an iron-based honeycomb lattice
  initiated by pressure-driven spin-crossover}},
\newblock Nat. Comm. \textbf{9}(1), 1914 (2018),
\newblock \doi{10.1038/s41467-018-04326-1}.

\bibitem{2020_Coak_TuneDimensionality_TMPS3}
M.~J. Coak, D.~M. Jarvis, H.~Hamidov, C.~R.~S. Haines, P.~L. Alireza, C.~Liu,
  S.~Son, I.~Hwang, G.~I. Lampronti, D.~Daisenberger, P.~Nahai-Williamson,
  A.~R. Wildes \emph{et~al.},
\newblock \emph{{Tuning dimensionality in van-der-Waals antiferromagnetic Mott
  insulators $TM$PS$_3$}},
\newblock J. Phys. Condens. Matter \textbf{32}(12), 124003 (2020),
\newblock \doi{10.1088/1361-648X/ab5be8}.

\bibitem{2022_Raman_DFT_FePS3}
S.~Das, S.~Chaturvedi, D.~Tripathy, S.~Grover, R.~Singh, D.~Muthu, S.~Sampath,
  U.~Waghmare and A.~Sood,
\newblock \emph{Raman and first-principles study of the pressure-induced
  mott-insulator to metal transition in bulk feps$_3$},
\newblock J. Phys. Chem. Solids \textbf{164}, 110607 (2022),
\newblock \doi{https://doi.org/10.1016/j.jpcs.2022.110607}.

\bibitem{2002_Rule_Contrasting_AFMorder_FePS3_MnPS3}
K.~Rule, S.~Kennedy, D.~Goossens, A.~Mulders and T.~Hicks,
\newblock \emph{{Contrasting antiferromagnetic order between FePS$_3$ and
  MnPS$_3$}},
\newblock Appl. Phys. A: Mater. Sci. Process. \textbf{74}(SUPPL.I), s811
  (2002),
\newblock \doi{10.1007/s003390201363}.

\bibitem{2016_Lancon_FePS3_MagStructure}
D.~Lan{\c{c}}on, H.~C. Walker, E.~Ressouche, B.~Ouladdiaf, K.~C. Rule, G.~J.
  McIntyre, T.~J. Hicks, H.~M. R{\o}nnow and A.~R. Wildes,
\newblock \emph{{Magnetic structure and magnon dynamics of the
  quasi-two-dimensional antiferromagnet FePS$_3$}},
\newblock Phys. Rev. B \textbf{94}(21), 1 (2016),
\newblock \doi{10.1103/PhysRevB.94.214407}.

\bibitem{1979_TMPX3_resistivity_bandGap}
R.~Brec, D.~M. Schleich, G.~Ouvrard, A.~Louisy and J.~Rouxel,
\newblock \emph{{Physical properties of lithium intercalation compounds of the
  layered transition-metal chalcogenophosphites}},
\newblock Inorg. Chem. \textbf{18}(7), 1814 (1979),
\newblock \doi{10.1021/ic50197a018}.

\bibitem{Zheng2019_abinito}
Y.~Zheng, X.~Jiang, X.~Xue, J.~Dai and Y.~Feng,
\newblock \emph{{Ab initio study of pressure-driven phase transition in
  FePS$_3$ and FePSe$_3$}},
\newblock Phys. Rev. B \textbf{100}(17), 174102 (2019),
\newblock \doi{10.1103/PhysRevB.100.174102}.

\bibitem{2020_Evarestov_DFT_FePX3}
R.~A. Evarestov and A.~Kuzmin,
\newblock \emph{{Origin of pressure‐induced insulator‐to‐metal transition
  in the van der Waals compound FePS$_3$ from first‐principles
  calculations}},
\newblock J. Comput. Chem. \textbf{41}(14), 1337 (2020),
\newblock \doi{10.1002/jcc.26178}.

\bibitem{2023_Jarvis_comparative_single_powder_FePS3}
D.~M. Jarvis, M.~J. Coak, H.~Hamidov, C.~R.~S. Haines, G.~I. Lampronti, C.~Liu,
  S.~Deng, D.~Daisenberger, D.~R. Allan, M.~R. Warren, A.~R. Wildes and S.~S.
  Saxena,
\newblock \emph{{Comparative structural evolution under pressure of powder and
  single crystals of the layered antiferromagnet FePS$_3$}},
\newblock Phys. Rev. B \textbf{107}, 54106 (2023),
\newblock \doi{10.1103/PhysRevB.107.054106}.

\bibitem{2007_Rule_neutron_MagneticStructure_FePS3}
K.~C. Rule, G.~J. McIntyre, S.~J. Kennedy and T.~J. Hicks,
\newblock \emph{{Single-crystal and powder neutron diffraction experiments on
  FePS$_3$: Search for the magnetic structure}},
\newblock Phys. Rev. B \textbf{76}(13), 134402 (2007),
\newblock \doi{10.1103/PhysRevB.76.134402}.

\bibitem{2020_Wildes_Biquadratic_FePS3}
A.~R. Wildes, M.~E. Zhitomirsky, T.~Ziman, D.~Lan{\c{c}}on and H.~C. Walker,
\newblock \emph{{Evidence for biquadratic exchange in the quasi-two-dimensional
  antiferromagnet FePS$_3$}},
\newblock J. Appl. Phys. \textbf{127}(22), 223903 (2020),
\newblock \doi{10.1063/5.0009114}.

\bibitem{2022_Zeng_FePS3_MagTran_DFT}
Y.~Zeng, D.-X. Yao and M.-R. Li,
\newblock \emph{{Ab initio study of magnetic structure transitions of
  ${\mathrm{FePS}}_{3}$ under high pressure}},
\newblock Phys. Rev. B \textbf{106}, 214408 (2022),
\newblock \doi{10.1103/PhysRevB.106.214408}.

\bibitem{2021_Coak_MagPhase_FePS3}
M.~J. Coak, D.~M. Jarvis, H.~Hamidov, A.~R. Wildes, J.~A.~M. Paddison, C.~Liu,
  C.~R.~S. Haines, N.~T. Dang, S.~E. Kichanov, B.~N. Savenko, S.~Lee,
  M.~Kratochv{\'{i}}lov{\'{a}} \emph{et~al.},
\newblock \emph{{Emergent Magnetic Phases in Pressure-Tuned van der Waals
  Antiferromagnet FePS$_3$}},
\newblock Phys. Rev. X \textbf{11}(1), 011024 (2021),
\newblock \doi{10.1103/PhysRevX.11.011024}.

\bibitem{2000_Ordering_SpinFlop_MnPS3}
D.~J. Goossens, A.~R. Wildes, C.~Ritter and T.~J. Hicks,
\newblock \emph{{Ordering and the nature of the spin flop phase transition in
  MnPS$_3$}},
\newblock J. Phys.: Condens. Matter \textbf{12}, 1845 (2000),
\newblock \doi{10.1088/0953-8984/12/8/327}.

\bibitem{2022_FM_monolayer_MP3}
C.~Autieri, G.~Cuono, C.~Noce, M.~Rybak, K.~M. Kotur, C.~E. Agrapidis,
  K.~Wohlfeld and M.~Birowska,
\newblock \emph{{Limited Ferromagnetic Interactions in Monolayers of $M$PS$_3$
  ($M$ = Mn and Ni)}},
\newblock J. Phys. Chem. C \textbf{126}(15), 6791 (2022),
\newblock \doi{10.1021/acs.jpcc.2c00646}.

\bibitem{2023_Field_induced_Spin_Reorientation_MnPS3}
H.~Han, H.~Lin, W.~Gan, R.~Xiao, Y.~Liu, J.~Ye, L.~Chen, W.~Wang, L.~Zhang,
  C.~Zhang and H.~Li,
\newblock \emph{{Field-induced spin reorientation in the N\'eel-type
  antiferromagnet ${\mathrm{MnPS}}_{3}$}},
\newblock Phys. Rev. B \textbf{107}, 075423 (2023),
\newblock \doi{10.1103/PhysRevB.107.075423}.

\bibitem{2000_CASTEP}
V.~Milman, B.~Winkler, J.~A. White, C.~J. Pickard, M.~C. Payne, E.~V.
  Akhmatskaya and R.~H. Nobes,
\newblock \emph{Electronic structure, properties, and phase stability of
  inorganic crystals: A pseudopotential plane-wave study},
\newblock Int. J. Quantum. Chem. \textbf{77}(5), 895 (2000),
\newblock \doi{10.1002/(SICI)1097-461X(2000)77:5<895::AID-QUA10>3.0.CO;2-C}.

\bibitem{2002_CASTEP}
M.~D. Segall, P.~J.~D. Lindan, M.~J. Probert, C.~J. Pickard, P.~J. Hasnip,
  S.~J. Clark and M.~C. Payne,
\newblock \emph{{First-principles simulation: ideas, illustrations and the
  CASTEP code}},
\newblock J. Phys. Condens. Matter \textbf{14}(11), 2717 (2002),
\newblock \doi{10.1088/0953-8984/14/11/301}.

\bibitem{Perdew1996GeneralizedSimple}
J.~P. Perdew, K.~Burke and M.~Ernzerhof,
\newblock \emph{{Generalized Gradient Approximation Made Simple}},
\newblock Phys. Rev. Lett. \textbf{77}(18), 3865 (1996),
\newblock \doi{10.1103/PhysRevLett.77.3865}.

\bibitem{pickard2006fly}
C.~J. Pickard,
\newblock \emph{On-the-fly pseudopotential generation in castep},
\newblock School of Physics and Astronomy, University of St Andrews St Andrews,
  KY16 9SS, United Kingdom  (2006).

\bibitem{1990_OTFP}
D.~Vanderbilt,
\newblock \emph{Soft self-consistent pseudopotentials in a generalized
  eigenvalue formalism},
\newblock Phys. Rev. B \textbf{41}, 7892 (1990),
\newblock \doi{10.1103/PhysRevB.41.7892}.

\bibitem{2014_DFT_error_estimate}
K.~Lejaeghere, V.~V. Speybroeck, G.~V. Oost and S.~Cottenier,
\newblock \emph{{Error Estimates for Solid-State Density-Functional Theory
  Predictions: An Overview by Means of the Ground-State Elemental Crystals}},
\newblock Crit. Rev. Solid State Mater. Sci. \textbf{39}(1), 1 (2014),
\newblock \doi{10.1080/10408436.2013.772503}.

\bibitem{1985_BFGS}
J.~D. Head and M.~C. Zerner,
\newblock \emph{{A Broyden-Fletcher-Goldfarb-Shanno optimization procedure for
  molecular geometries}},
\newblock Chem. Phys. Lett. \textbf{122}(3), 264 (1985),
\newblock \doi{10.1016/0009-2614(85)80574-1}.

\bibitem{2009_TS_correction}
A.~Tkatchenko and M.~Scheffler,
\newblock \emph{{Accurate Molecular Van Der Waals Interactions from
  Ground-State Electron Density and Free-Atom Reference Data}},
\newblock Phys. Rev. Lett. \textbf{102}(7), 073005 (2009),
\newblock \doi{10.1103/PhysRevLett.102.073005}.

\bibitem{1963_Hubbard_elelctron_correlations}
J.~Hubbard,
\newblock \emph{Electron correlations in narrow energy bands},
\newblock Proc. R. Soc. A. \textbf{276}(1365), 238 (1963),
\newblock \doi{10.1098/rspa.1964.0019}.

\bibitem{1991_DFT+U}
V.~I. Anisimov, J.~Zaanen and O.~K. Andersen,
\newblock \emph{{Band theory and Mott insulators: Hubbard \textit{U} instead of
  Stoner \textit{I}}},
\newblock Phys. Rev. B \textbf{44}(3), 943 (1991),
\newblock \doi{10.1103/PhysRevB.44.943}.

\bibitem{1995_MottHubbard_Insulators}
A.~I. Liechtenstein, V.~I. Anisimov and J.~Zaanen,
\newblock \emph{{Density-functional theory and strong interactions: Orbital
  ordering in Mott-Hubbard insulators}},
\newblock Phys. Rev. B \textbf{52}(8), R5467 (1995),
\newblock \doi{10.1103/PhysRevB.52.R5467}.

\bibitem{2008_lebegue_DFT+U_Fe}
S.~Leb{\`e}gue, S.~Pillet and J.~{\'A}ngy{\'a}n,
\newblock \emph{{Modeling spin-crossover compounds by periodic DFT+\textit{U}
  approach}},
\newblock Phys. Rev. B \textbf{78}(2), 024433 (2008),
\newblock \doi{10.1103/PhysRevB.78.024433}.

\bibitem{1977_MP_kmesh}
J.~D. Pack and H.~J. Monkhorst,
\newblock \emph{{"Special points for Brillouin-zone integrations"---a reply}},
\newblock Phys. Rev. B \textbf{16}, 1748 (1977),
\newblock \doi{10.1103/PhysRevB.16.1748}.

\bibitem{Pickard2011AbSearching}
C.~J. Pickard and R.~J. Needs,
\newblock \emph{{Ab initio random structure searching}},
\newblock J. Phys. Cond. Matter \textbf{23}(5) (2011),
\newblock \doi{10.1088/0953-8984/23/5/053201}.

\bibitem{Kennedy2002_SWARM}
J.~Kennedy and R.~Eberhart,
\newblock \emph{A discrete binary version of the particle swarm algorithm},
\newblock In \emph{1997 IEEE International Conference on Systems, Man, and
  Cybernetics. Computational Cybernetics and Simulation}, vol.~5, pp.
  4104--4108 vol.5,
\newblock \doi{10.1109/ICSMC.1997.637339} (1997).

\bibitem{wang2012calypso}
Y.~Wang, J.~Lv, L.~Zhu and Y.~Ma,
\newblock \emph{Calypso: A method for crystal structure prediction},
\newblock Comput. Phys. Commun. \textbf{183}(10), 2063 (2012),
\newblock \doi{10.1016/j.cpc.2012.05.008}.

\bibitem{wu2013adaptive_genetic}
S.~Wu, M.~Ji, C.-Z. Wang, M.~C. Nguyen, X.~Zhao, K.~Umemoto, R.~Wentzcovitch
  and K.-M. Ho,
\newblock \emph{An adaptive genetic algorithm for crystal structure
  prediction},
\newblock J. Condens. Matter Phys. \textbf{26}(3), 035402 (2013),
\newblock \doi{10.1088/0953-8984/26/3/035402}.

\bibitem{2021_Ma_NiPS3}
X.~Ma, Y.~Wang, Y.~Yin, B.~Yue, J.~Dai, J.~Cheng, J.~Ji, F.~Jin, F.~Hong, J.~T.
  Wang, Q.~Zhang and X.~Yu,
\newblock \emph{{Dimensional crossover tuned by pressure in layered magnetic
  NiPS$_3$}},
\newblock Sci. China: Phys. Mech. Astron. \textbf{64}, 1 (2021),
\newblock \doi{10.1007/S11433-021-1727-6}.

\bibitem{klingen1973_FePX3_structure}
W.~Klingen, G.~Eulenberger and H.~Hahn,
\newblock \emph{{Uber die kristallstrukturen von Fe$_2$P$_2$Se$_6$ und
  Fe$_2$P$_2$S$_6$}},
\newblock Z. fur Anorg. Allg. Chem. \textbf{401}(1), 97 (1973),
\newblock \doi{10.1002/zaac.19734010113}.

\bibitem{Ouvrard1985_TMPS3_structures}
G.~Ouvrard, R.~Brec and J.~Rouxel,
\newblock \emph{{Structural determination of some MPS$_3$ layered phases (M =
  Mn, Fe, Co, Ni and Cd)}},
\newblock Mater. Res. Bull. \textbf{20}(10), 1181  (1985),
\newblock \doi{10.1016/0025-5408(85)90092-3}.

\bibitem{2020_Neal_TMPS3_symmCrossover_Layers}
S.~N. Neal, H.~S. Kim, A.~V. O'neal, K. R.and~Haglund, D.~G. Smith, K.
  A.and~Mandrus, H.~A. Bechtel, G.~L. Carr, K.~Haule, D.~Vanderbilt and J.~L.
  Musfeldt,
\newblock \emph{{Symmetry crossover in layered $M$PS$_3$ complexes ($M$=Mn, Fe,
  Ni) via near-field infrared spectroscopy}},
\newblock Phys. Rev. B \textbf{102}, 85408 (2020),
\newblock \doi{10.1103/PhysRevB.102.085408}.

\bibitem{PhysRevLett.78.4063}
K.~Parlinski, Z.~Q. Li and Y.~Kawazoe,
\newblock \emph{First-principles determination of the soft mode in cubic
  ${\mathrm{zro}}_{2}$},
\newblock Phys. Rev. Lett. \textbf{78}, 4063 (1997),
\newblock \doi{10.1103/PhysRevLett.78.4063}.

\bibitem{PhysRevB.92.184301}
J.~H. Lloyd-Williams and B.~Monserrat,
\newblock \emph{Lattice dynamics and electron-phonon coupling calculations
  using nondiagonal supercells},
\newblock Phys. Rev. B \textbf{92}, 184301 (2015),
\newblock \doi{10.1103/PhysRevB.92.184301}.

\bibitem{2012_nicholls_optados}
R.~J. Nicholls, A.~J. Morris, C.~J. Pickard and J.~R. Yates,
\newblock \emph{{OptaDOS} - a new tool for {EELS} calculations},
\newblock J. Phys. Conf. \textbf{371}, 012062 (2012),
\newblock \doi{10.1088/1742-6596/371/1/012062}.

\bibitem{2014_morris_optados}
A.~J. Morris, R.~J. Nicholls, C.~J. Pickard and J.~R. Yates,
\newblock \emph{Optados: A tool for obtaining density of states, core-level and
  optical spectra from electronic structure codes},
\newblock Comput. Phys. Commun. \textbf{185}(5), 1477 (2014),
\newblock \doi{10.1016/j.cpc.2014.02.013}.

\bibitem{2012_Wildes_magnon_SpinExchange_FePS3}
A.~R. Wildes, K.~C. Rule, R.~I. Bewley, M.~Enderle and T.~J. Hicks,
\newblock \emph{{The magnon dynamics and spin exchange parameters of
  FePS$_3$}},
\newblock J. Phys. Condens. Matter \textbf{24}(41), 416004 (2012),
\newblock \doi{10.1088/0953-8984/24/41/416004}.

\bibitem{1986_Grasso_optical_TMPS3}
V.~Grasso, S.~Santangelo and M.~Piacentini,
\newblock \emph{{Optical absorption spectra of some transition metal
  thiophosphates}},
\newblock Solid State Ion. \textbf{20}(1), 9 (1986),
\newblock \doi{10.1016/0167-2738(86)90028-7}.

\bibitem{2006_kotliar_electronic_DMFT}
G.~Kotliar, S.~Y. Savrasov, K.~Haule, V.~S. Oudovenko, O.~Parcollet and
  C.~Marianetti,
\newblock \emph{Electronic structure calculations with dynamical mean-field
  theory},
\newblock Rev. Mod. Phys. \textbf{78}(3), 865 (2006),
\newblock \doi{10.1103/RevModPhys.78.865}.

\bibitem{2006_grimme_semiempirical}
S.~Grimme,
\newblock \emph{{Semiempirical GGA-type density functional constructed with a
  long-range dispersion correction}},
\newblock J. Comput. Chem. \textbf{27}(15), 1787 (2006),
\newblock \doi{10.1002/jcc.20495}.

\bibitem{2008_VASP_DFT}
J.~Hafner,
\newblock \emph{{Ab-initio simulations of materials using VASP:
  Density-functional theory and beyond}},
\newblock J. Comput. Chem. \textbf{29}(13), 2044 (2008),
\newblock \doi{10.1002/jcc.21057}.

\bibitem{2008_perdew_PBESol}
J.~P. Perdew, A.~Ruzsinszky, G.~I. Csonka, O.~A. Vydrov, G.~E. Scuseria, L.~A.
  Constantin, X.~Zhou and K.~Burke,
\newblock \emph{Restoring the density-gradient expansion for exchange in solids
  and surfaces},
\newblock Phys. Rev. Lett. \textbf{100}(13), 136406 (2008),
\newblock \doi{10.1103/PhysRevLett.100.136406}.

\end{thebibliography}

\nolinenumbers
\end{document}